\documentclass{ajour}

\usepackage{amsmath}
\usepackage{amsfonts}
\usepackage{epsfig}
\newcommand{\bea}{\begin{eqnarray}}
\newcommand{\eea}{\end{eqnarray}}
\newcommand{\hs}[1]{\hspace*{#1cm}}
\newcommand{\vs}[1]{\vspace*{#1cm}}
\newcommand{\la}{\langle}
\newcommand{\ra}{\rangle}
\newcommand{\half}{{\textstyle \frac{1}{2}}}
\newcommand{\ora}[1]{\overrightarrow{#1}}
\newcommand{\lra}{\!\leftrightarrow\!}
\newcommand{\balpha}{\mbox{\boldmath $\alpha$}}

\newcommand{\bbeta}{\mbox{\boldmath $\beta$}}
\newcommand{\bgamma}{\mbox{\boldmath $\gamma$}}
\newcommand{\bepsilon}{\mbox{\boldmath $\epsilon$}}
\newcommand{\bLambda}{\mbox{\boldmath $\Lambda$}}

\newcommand{\bp}{\mbox{\boldmath $p$}}

\newcommand{\bx}{\mbox{\boldmath $x$}}
\newcommand{\by}{\mbox{\boldmath $y$}}
\newcommand{\bA}{\mbox{\boldmath $A$}}

\newcommand{\bH}{\mbox{\boldmath $H$}}
\newcommand{\bI}{\mbox{\boldmath $I$}}
\newcommand{\bN}{\mbox{\boldmath $N$}}

\newcommand{\bS}{\mbox{\boldmath $S$}}

\newcommand{\bV}{\mbox{\boldmath $V$}}
\newcommand{\clf}{\mathcal{F}}
\newcommand{\clh}{\mathcal{H}}

\newcommand{\cll}{\mathcal{L}}
\newcommand{\clp}{\mathcal{P}}
\newcommand{\grad}{\nabla}
\def\tightmaths{                                                                
  \thinmuskip=1.5mu                                                             
  \medmuskip=2mu plus 1mu minus 2mu                                             
  \thickmuskip=2.5mu plus 2.5mu
}                         
                                  
\tightmaths

\begin{document}

%


\authorrunninghead{C. E. Dolby and S. F. Gull}
\titlerunninghead{Arbitrary Observers In Electromagnetic Backgrounds}





\title{New Approach To Quantum Field Theory For Arbitrary Observers In Electromagnetic Backgrounds}


\authors{Carl E. Dolby and Stephen F. Gull}
\affil{Astrophysics Group, Cavendish Laboratory, Madingley Road, Cambridge
CB3 0HE, U.K.}

\email{c.dolby@mrao.cam.ac.uk}

\abstract{A reformulation of fermionic QFT in electromagnetic backgrounds is 
presented which uses methods analogous to those of conventional 
multiparticle quantum mechanics. Emphasis is placed on the (Schr\"{o}dinger picture) 
states of the system, described in terms of Slater
determinants of Dirac states, and not on the field operator $\hat{\psi}(x)$ (which 
is superfluous in this
approach). The vacuum state `at time $\tau$' is defined as
the Slater determinant of a basis for the span of the
negative spectrum of the `first quantized' Hamiltonian $\hat{H}(\tau)$, thus providing 
a concrete realisation of the Dirac Sea. The general S-matrix element of
the theory is derived in terms of time-dependent
Bogoliubov coefficients, demonstrating that the S-matrix follows
directly from the definition of inner product between Slater
determinants. The process of `Hermitian extension', inherited directly from 
conventional multiparticle quantum mechanics, allows second 
quantized operators to be defined without appealing to a 
complete set of orthonormal modes, and provides an extremely 
straightforward derivation of the 
general expectation value of the theory. The 
concept of `radar time', advocated by Bondi in his work on k-calculus, 
is used to  
generalise the particle interpretation to an arbitrarily moving observer. A 
definition of particle results, which depends {\it only} on the observer's motion and the 
background present, not on any choice of coordinates or gauge, or of the 
particle detector. We relate this approach 
to conventional methods by comparing and contrasting various derivations. 
Our particle definition can be viewed as a 
generalisation to arbitrary observers of Gibbons' approach \cite{Gibb2}.}


\begin{article}

\section{INTRODUCTION}

We present here an initial-value formulation of 
fermionic QFT in an electromagnetic background. This formulation can be seen as the natural relativistic 
generalisation of non-relativistic multiparticle quantum 
mechanics. We emphasise the 
states of the system, described in terms of Slater
determinants of Dirac states, rather than the field operator. The vacuum is defined as
the Slater determinant of a basis for the span of the
negative spectrum of the `first quantized' Hamiltonian, providing  
a concrete manifestation of the Dirac Sea. Simple derivations of the general S-Matrix element and expectation values in the theory then follow. Moreover, 
this approach suggests a consistent particle interpretation at all times, without 
requiring any `asymptotic niceness conditions' on the `in' and `out' states. By using the 
concept of `radar time' (originally made popular by Bondi in his work on k-calculus
\cite{Bohm}) we 
generalise this particle interpretation to an arbitrarily moving observer, providing a 
definition of particle which depends {\it only} on the observer's motion and on the 
background.

	Perhaps the most surprising aspect of this description of relativistic multiparticle
systems is that it is essentially new. Even in 1932, when the `Dirac Sea' 
concept was first introduced \cite{Dirac3,Dirac4}, the notion of using 
Slater determinants to describe the `filling of energy levels' was 
well-known \cite{Slater}. Indeed, a formulation of relativistic fermionic 
systems close to that presented here (although not in a 
classical background, not explicitly using Slater determinants, and not considering 
observer dependence) was  
presented in 1934 by Furry and Oppenheimer \cite{FuOp} (it is described from a more 
modern standpoint in the introduction to Weinberg's book \cite{Wein}). However, the 
view at the time was that the Dirac Sea was little more than a  
mathematical trick, and should be 
stable and trivial to describe. Even Dirac \cite{Dirac} 
stated that ``The vacuum is quite a trivial thing physically, and we 
should expect it to correspond to a trivial solution of the Schr\"{o}dinger 
equation''. The Furry-Oppenheimer theory was rapidly replaced by the 
Canonical approach, which treats the `holes' in the Dirac 
Sea (rather than its constituents) on the same footing as 
the electrons, while defining the vacuum only implicitly, as `containing no electrons or holes'.

	A more recent formulation of fermionic QFT in classical backgrounds, which 
also actively incorporates the Dirac Sea concept, has been given by  
Keifer et al. \cite{Keif1,Keif2,Keif3}, Floreanini et al. \cite{Flor} 
and others. This is based on the `functional Schr\"{o}dinger equation', where the 
state of the system is described by a functional of Grassmann fields, 
much as was proposed by Berezin \cite{Ber}. 
This formulation has met with considerable success in applications \cite{Keif2,Keif3,Hal}, 
although questions regarding particle interpretation, the choice of 
boundary conditions, and the foliation of spacetime remain unresolved.

	The formulation of QFT presented in the present paper resolves 
these problems. We support the claim \cite{Jackiw} that, when considering QFT in 
electromagnetic or gravitational backgrounds (so that vacuum effects are important) 
 it is clearer and faster to work directly with a concrete representation of the Dirac sea. Since this approach is the natural relativistic generalisation of methods 
already familiar in conventional multiparticle quantum mechanics, we also believe 
that it will often provide a favourable alternative to the canonical approach.

	Sections 2 and 3 set out  
the basis of our approach, as described in the rest frame of an inertial observer. The 
State space $\clf_{\wedge}(\clh)$ is presented in Section 2 in terms of Slater 
determinants of spinor-valued functions. In Section 3 we describe the time-dependent particle
	interpretation of state space. We show how the general S-Matrix element of the theory follows directly 
from the definition of inner product between Slater determinants, and we calculate the general 
expectation value using the theory. 
`Radar time' is introduced in Section 4, and is used to 
generalise the particle interpretation to arbitrarily 
moving observers.
In Section 5 
we show  
how our formalism is related to more conventional methods, using the 
field operator $\hat{\psi}(x)$. By extending a technique originally 
developed for real scalar fields by DeWitt \cite{DeWitt}, we rederive the 
general S-Matrix element of the theory and compare this to the derivation 
presented in Section 3. We also show how our particle definition 
can be expressed in more conventional terms by discussing it's relation to the method of Hamiltonian diagonalisation. Our definition consistently combines the conventional `Bogoliubov coefficient
	method' with the `tunnelling method', resolving the gauge
	inconsistencies \cite{SPad} that trouble each of these methods. A retrospective overview is presented in Section 6.

\section{THE STATE SPACE}

\subsection{Preliminaries}

The Lagrangian density for the Dirac equation in an electromagnetic
background $A_{\mu}(x)$ is \cite{Ryder}
\begin{equation} \cll = \bar{\psi}(x)(i \gamma^{\mu}\overleftrightarrow{\grad}_{\mu} - m) 
\psi(x) = \frac{i}{2} [\bar{\psi} \gamma^{\mu} \grad_{\mu} \psi - \grad_{\mu} \bar{\psi} 
\gamma^{\mu} \psi ] - m \bar{\psi} \psi
 \label{eq:diss2.1}\end{equation}
where $\{ \gamma^{\mu} , \gamma^{\nu} \} = 2 \eta^{\mu \nu} I_4$, ($I_4$ is 
the $4 \times 4$ identity matrix and \linebreak $\eta^{\mu \nu} = {\rm diag}(1,-1,-1,-1)$), $\psi(x)$ 
is a 4-component spinor, $\bar{\psi}(x) \equiv
	\psi^{\dagger}(x) \gamma^0$, and $e$ is the charge of the
	fermion ($e < 0$ for electrons). The {\it covariant derivative} $\grad_{\mu}$ is 
defined by $\grad_{\mu} \psi(x) \equiv \partial_{\mu} \psi(x) + i e A_{\mu}(x)  \psi(x)$.
	The Lagrangian (\ref{eq:diss2.1}) leads to the governing
	equation:
\begin{equation} (i \gamma^{\mu} \grad_{\mu} - m) \psi(x) = 0 \label{eq:diss2.2}\end{equation}

with a conserved inner product:
\begin{equation} \la \psi(\bx,t) | \phi(\bx,t) \ra = \int {\rm d}^3 \bx \ \psi^{\dagger}(\bx,t) 
\phi(\bx,t) \label{eq:diss2.6}\end{equation}

 A `first quantized' Dirac state is a spinor valued
function of space, which evolves with time. The `first quantized'
state space $\clh$ is the set of all spinor-valued functions of
$\mathbb{R}^3$ whose norm under (\ref{eq:diss2.6}) is finite,
$\clh \equiv L^2(\mathbb{R}^3)^4$. We shall denote a first quantized state 
at time $t$ by $\psi(\bx,t)$ or $|\psi(t) \ra$ or, where no ambiguity is possibly, simply $\psi$ (there will be little need to distinguish between a state and its coordinate representation).

	It is convenient to write (\ref{eq:diss2.2}) in  `Hamiltonian form' as:
\begin{align} i \grad_0 \psi(\bx,t) & =  \hat{H}_1(\bx,t) \psi(\bx,t) \label{eq:diss2.4} \\
\mbox{ where } \hat{H}_1(\bx,t) \psi(\bx,t) & = (- i \gamma^0 \gamma^k \grad_k 
+ m \gamma^0 ) \psi(\bx,t) \label{eq:diss2.3} \end{align}
	and $k$ is to be summed over $k = 1,2,3$. This defines the
	(gauge covariant) {\it first quantized Hamiltonian operator}
	$\hat{H}_1(\bx,t)$, which plays an important role below.
	The expectation value of
	$\hat{H}_1(\bx,t)$ in the state $\psi(\bx,t)$ is:
\begin{align} \la \psi(\bx,t) | \hat{H}_1(\bx,t) | \psi(\bx,t) \ra  & = \hs{-.1}  
\int {\rm d}^3 \bx \ \bar{\psi}(\bx,t) (- i \gamma^{k} \grad_k \hs{-.1} + \hs{-.1} m ) \psi(\bx,t) \label{blobby} \\
& =  \hs{-.1} \int {\rm d}^3 \bx \ T^{0}_{\hs{.2} 0, \psi}(x) \label{eq:enmom2} \hs{.2} 
\equiv H_{t_0}(\psi) \\
\mbox{ where } T^{\mu}_{\hs{.2} \nu,\psi}(x) & = i \bar{\psi} \gamma^{\mu} 
\overleftrightarrow{\grad}_{\nu} \psi - \delta^{\mu}_{\nu} \cll \label{eq:enmom} \end{align}
	is the energy-momentum tensor. The expectation value of $\hat{H}_1$ is 
identified as the 
spatial integral of the `00-component' of the energy-momentum tensor. If 
we require that $\psi(\bx,t)$ is a solution of the Dirac equation, 
the $\delta^{\mu}_{\nu} \cll$ term and the $\overleftrightarrow{}$ in (\ref{eq:enmom}) both vanish 
and we can write $T^0_{\hs{.2} 0}(x) = i \bar{\psi} \gamma^0 \grad_0 \psi$. Also notice that if the 
LHS of (\ref{eq:diss2.4}) had been $i \frac{\partial
	\psi}{\partial t}$ rather than $i
	\grad_0 \psi = i \frac{\partial \psi}{\partial t} - e A_{0}(x)
	\psi$, then the RHS would have been
	$\hat{H}_{ev} = \hat{H}_1 + e A_0(x)$, which is
	clearly dependent on gauge.

\subsection{The Full Fock Space Over $\clh$}

The {\it antisymmetric Fock Hilbert space} over the complex Hilbert
space $\clh$ is denoted $\clf_{\wedge}(\clh)$ and is defined \cite{Ott} in
terms of the {\it antisymmetric Tensor Algebra} over $\clh$. It is 
a natural and familiar construction by which a
 quantum theory of fermions can be formulated. We now define $\clf_{\wedge}(\clh)$. Let $\clh$
be the Hilbert space in the previous Section, with inner product denoted by 
$\la \hs{.2}| \hs{.2} \ra$. Let $\otimes^n \clh$ denote the direct product of 
$n$ copies of
$\clh$, and let $\wedge^n \clh$ denote the restriction of $\otimes^n
\clh$ to those `states' which are completely antisymmetric under
changes in the order of the elements $|\psi\ra \in \clh$ from which it is
constructed. Given $|\psi_1\ra |\psi_2\ra \dots |\psi_n \ra 
\in
\otimes^n \clh$ we can define $\psi_1 \wedge \psi_2 \wedge \dots 
\wedge \psi_n \in \wedge^n \clh$ by:
\begin{equation}\psi_1 \wedge \psi_2 \wedge \dots 
\wedge \psi_n \equiv 
\frac{1}{\sqrt{n!}} \sum_{\sigma} {\rm sign}(\sigma) |\psi_{\sigma(1)}\ra 
|\psi_{\sigma(2)}\ra \dots |\psi_{\sigma(n)}\ra \label{eq:oplus}\end{equation}
	where $\{ \sigma(i), i=1,\dots ,n \}$ is a permutation of 
$\{ 1 \dots n \}$. This is simply the Slater determinant of the states 
$|\psi_1\ra \dots |\psi_n\ra$. The antisymmetric Fock Hilbert space $\clf_{\wedge}(\clh)$ is now given by:
\begin{equation} \clf_{\wedge}(\clh) = \oplus_{n=0}^{\infty} \wedge^n 
\clh \notag \end{equation}
	where $\wedge^0 \clh \equiv \mathbb{C}$ and $\wedge^1 \clh
	\equiv \clh$. States which lie entirely within $\wedge^r \clh$
	for some $r$ are said to
	be of {\it grade} r. 

A useful operation on $\clf_{\wedge}(\clh)$ is the
`inner derivative' (named by analogy with differential
geometry)  $i_{\psi} : \wedge^n \clh
\rightarrow \wedge^{n-1} \clh$. This is defined by:
\begin{equation} i_{\psi}: \phi_1 \wedge \dots \wedge \phi_n \rightarrow 
 \sum_i (-)^{i+1} \la \psi | \phi_i \ra \phi_1 \wedge
 \dots \wedge \check{\phi}_i \wedge \dots \wedge \phi_n 
\label{eq:dot}\end{equation} 
	 where the check over $\phi_i$ signifies that this state is omitted 
from the product. The relation $i_{\psi} : \clf_{\wedge}(\clh)
\rightarrow \clf_{\wedge}(\clh)$ is obtained from (\ref{eq:dot}) by imposing 
linearity, together with the additional convention $i_{\psi} \lambda = 0$ for 
$\lambda \in \wedge^0 \clh$. It is clear that $i_{\psi} (i_{\psi} | F \ra) = 0$ for all 
$| F \ra \in \clf_{\wedge}(\clh)$, and that:
\begin{equation} i_{\psi} (\phi \wedge |F \ra) = \la \psi | \phi \ra |F \ra -
\phi \wedge (i_{\psi} | F \ra) \label{eq:CARprim} \end{equation}
The operation $i_{\psi}$ is denoted as $a(\psi)$ by Ottesen \cite{Ott}, and plays the
role of an annihilation operator. Here $i_{\psi}$ will play a
similar, although not identical role.

	Finally, the inner product on 
$\clf_{\wedge}(\clh)$ is given by: 
\begin{equation} \la \psi_1 \wedge \dots \wedge \psi_n | 
\phi_1 \wedge \dots \wedge \phi_m \ra = \delta_{n m} 
\det [ \la \psi_i | \phi_j \ra ] \label{eq:inprod} \end{equation}
	where $\la \psi_i | \phi_j \ra$ refers to the inner 
	product on $\clh$. (For states $\lambda,\mu \in \wedge^0 \clh$ define 
$\la \lambda | \mu \ra = \bar{\lambda} \mu$ and $\la \lambda | F_n \ra = 0$ for any 
state $|F_n\ra$ of grade $n > 0$.) This agrees with the inner
	product defined in terms of Slater determinants. 
Although we use the notation $\la \hs{.2} | \hs{.2} \ra$ to refer to both the inner product 
on $\clh$ and the inner product on $\clf_{\wedge}(\clh)$, 
it will be clear from the context which is meant.

\subsection{Operators on Fock Space}

Let $\hat{A}_1:\clh \rightarrow \clh$ be an operator on the space of
Dirac states.  We wish to construct from it an operator which can act
on all of state space. There are two useful ways of doing this: {\it Hermitian extension} $\hat{A}_H: \clf_{\wedge}(\clh)
\rightarrow \clf_{\wedge}(\clh)$, and {\it Unitary extension}
$\hat{A}_U: \clf_{\wedge}(\clh) \rightarrow \clf_{\wedge}(\clh)$ (outlined also in
Ottesen \cite{Ott}). These are defined by:

\begin{align} \hat{A}_H : \psi_1 \wedge \psi_2 \wedge \dots 
\wedge \psi_N  & \rightarrow  \sum_{i=1}^{N} \psi_1 \wedge \dots (\hat{A}_1 \psi_i) \wedge 
\psi_{i+1} \dots \wedge \psi_N \label{eq:SSB8} \\
 \hat{A}_U : \psi_1 \wedge \psi_2 \wedge \dots 
\wedge \psi_N & \rightarrow (\hat{A}_1\psi_1) \wedge (\hat{A}_1\psi_2) 
\wedge \dots \wedge (\hat{A}_1\psi_N) \label{eq:SSB9}\end{align}

If $\hat{A}_1$ is (anti)hermitian
with respect to the inner product (\ref{eq:diss2.6}) on $\clh$, then $\hat{A}_H$ is (anti)hermitian
with respect to the inner product (\ref{eq:inprod}) on $\clf_{\wedge}(\clh)$. If $\hat{A}_1$ is unitary,
then so is $\hat{A}_U$. Also $(e^{\hat{A}_1})_U =
e^{\hat{A}_H}$, so that if $\hat{U}_1 = e^{\hat{A}_1}$ on $\clh$ then $\hat{U}_U = e^{\hat{A}_H}$ on
$\clf_{\wedge}(\clh)$.

\subsubsection*{Some Simple Properties}

\begin{enumerate}

\item $(\hat{A} + \hat{B})_H = \hat{A}_H + \hat{B}_H$, $[ \hat{A}_H,\hat{B}_H ] = [ \hat{A} , \hat{B} ]_H$ and $(\hat{A}
\hat{B})_U = \hat{A}_U \hat{B}_U$.

\item $[ \hat{A}_H , \psi \wedge ] = (\hat{A}_1 \psi) $ and $[ \hat{A}_H , i_{\psi} ] = - i_{\hat{A}^{\dagger}_1 \psi}$ 

\item If $ \psi_1, \psi_2, \dots \psi_N $ are all 
eigenstates of $\hat{A}_1$
with eigenvalues $\lambda_1, \dots \lambda_N$, then $\psi_1 \wedge 
\psi_2 \dots \wedge \psi_N$ is an eigenstate of $\hat{A}_H$ with 
eigenvalue $\sum_{i = 1}^{N} \lambda_i$.

\item If  $ \psi_1, \psi_2, \dots \psi_N $ are orthonormal 
and $| F \ra \equiv | \psi_1 \wedge \psi_2 \dots \wedge \psi_N \ra$
then 
\begin{align} \la F | \hat{A}_H | F \ra
 & = \sum_{i = 1}^{N} \la \psi_i | \hat{A}_1 | \psi_i \ra \label{eq:prop3a}\\
\la F |(\hat{A}_H)^2 | F \ra 
& = \sum_{i = 1}^{N} \la \psi_i | \hat{A}_1^2 | \psi_i \ra  + 
 2 \sum_{i < j} ( \la \psi_i | \hat{A}_1 | \psi_i \ra \la \psi_j | \hat{A}_1 
| \psi_j \ra \label{eq:prop3b} \\ 
& \hs{4} - \la \psi_i | \hat{A}_1 | \psi_j \ra \la \psi_j | \hat{A}_1 
| \psi_i \ra) \notag \\
 \la F |(\hat{A}_{H})^2 | F \ra & - (\la F | \hat{A}_{H} | F \ra)^2 = 
\sum_i  \la \psi_i | \hat{A}_1^2 | \psi_i \ra  -  
\sum_{i,j} | \la \psi_i |\hat{A}_1 |\psi_j \ra |^2 \label{eq:prop3c} \end{align}

\end{enumerate}

\subsection{Evolution of States}
 
  Define the evolution operator 
$\hat{U}_1(t,t_0)$ on $\clh$ by
\begin{equation} \hat{U}_1(t,t_0) |\psi_{t_0} \ra \equiv |\psi_{t_0}(t)\ra 
\label{eq:SSB25}\end{equation}
 where $|\psi_{t_0}\ra$ represents the chosen initial conditions
 $\psi_{t_0}(\bx)$ at time $t_0$, and $|\psi_{t_0}(t)\ra$ represents the
 solution $\psi_{t_0}(\bx,t)$ of the Dirac equation satisfying these
 initial conditions.

	We consider only QFT in an (external) electromagnetic background, 
so that we ignore direct particle-particle interactions and just work within 
the `zeroth order Hartree Fock' approximation. This assumes that the 
evolution operator on $\clf_{\wedge}(\clh)$ is just the unitary extension of 
the evolution operator on $\clh$, or that $\hat{H}(t) = \hat{H}_{1,H}(t)$, which is the 
natural generalisation of the equation that appears in multiparticle quantum 
mechanics textbooks \cite{Szab,MYS} as $\hat{H}(\bx_1, \dots \bx_n) = \hat{H}(\bx_1) 
+ \dots + \hat{H}(\bx_n)$. The action of $\hat{U}(t,t_0)$ is given by:
\begin{equation} \hat{U}(t,t_0): \psi_{1,t_0} \wedge \dots \wedge \psi_{n,t_0} 
\rightarrow  \psi_{1,t_0}(t) \wedge \dots \wedge \psi_{n,t_0}(t) \label{eq:evol} \end{equation}
The multiparticle solution is simply the Slater 
determinant of the appropriate `first quantized' solutions. This construction preserves grade, and 
implies that the unitarity of 
$\hat{U}(t,t_0)$ follows immediately from the unitarity of the first 
quantized 
Dirac equation.

We have now a state space, an evolution equation (\ref{eq:evol}) and a 
conserved inner product (\ref{eq:inprod}). This is all we need to calculate 
arbitrary S-Matrix elements (from (\ref{eq:evol}) 
and (\ref{eq:inprod})), arbitrary expectation values (from (\ref{eq:prop3a})), 
and even fluctuations in these expectation 
values (from (\ref{eq:prop3c})). However,  
the theory is not invested with physical meaning 
until the states of the system can be specified in terms of their physical 
properties. Accordingly we now turn our attention 
to a particle interpretation.

\section{A PARTICLE INTERPRETATION OF STATE SPACE}

\subsection{The Positive/Negative Energy Split and the Vacuum State}

 Consider the action of $\hat{H}_1(t_0)$ on $\clh$, at some fixed time
	$t_0$. Since $\hat{H}_1(t_0)$ is Hermitian we can
	parametrise $\clh$ in terms of the eigenvectors of
	$\hat{H}_1(t_0)$. From this we can define $\clh^{\pm}(t_0)$ such that:
\begin{align} \clh^+(t_0) & \mbox{ is the span of the positive spectrum of } 
	\hat{H}_1(t_0) \notag \\
\clh^-(t_0) & \mbox{ is the span of the negative spectrum of }
	\hat{H}_1(t_0) \notag \end{align}
$\clh^+(t_0)$ is the set of all
	positive energy states, and $\clh^-(t_0)$ is the set of all
	negative energy states as defined at time $t_0$. $\clh^{\pm}(t_0)$ can alternatively be 
defined such that the projection operators $\hat{P}^{\pm}(t_0): \clh
\rightarrow \clh^{\pm}(t_0)$  are orthogonal projections
satisfying:
\begin{equation} H_{t_0}(\hat{P}^+(t_0) \psi) \geq H_{t_0}(\psi) \geq 
H_{t_0}(\hat{P}^-(t_0) \psi) \label{eq:Hdef} \end{equation}
(in the notation of (\ref{eq:enmom2})) for all $\psi \in \clh$. These definitions are equivalent (as shown in \cite{mythesis}) but
	since (\ref{eq:Hdef}) does not refer explicitly to the spectrum of
	$\hat{H}_1(t_0)$ it is often more useful, as in section 4.1. Also, we 
have implicitly assumed that $\hat{H}_1(t)$ has no zero energy eigenstates (for 
any $t$). This is true for a large class of observers, but no longer holds when particle horizons are present. In the presence of particle horizons we have a third space $\clh^0(t)$, consisting of zero energy 
eigenstates. This case is explained in detail in \cite{Me2}; the 
simplest example is the Unruh effect, which will be  
discussed briefly in Section 4.1.

	We can now
	define the {\it vacuum at time $t_0$}, $| {\rm vac}_{t_0} \ra$ as follows:
\begin{align} | {\rm vac}_{t_0} \ra & \mbox{ is the Slater determinant of any basis of }
	\clh^-(t_0), \notag \\
& \mbox{ normalised so that } \la {\rm vac}_{t_0} | {\rm vac}_{t_0} \ra = 1. \notag \end{align}
	This specifies
	$| {\rm vac}_{t_0} \ra$ up to an arbitrary phase factor. It is the state 
	in which all negative energy degrees of freedom are full, and hence 
	is a concrete manifestation of the Dirac Sea.

	To illustrate this, suppose temporarily that $\clh$ contains only $N$ 
positive and $N$ negative energy degrees of freedom ($N \rightarrow \infty$ contains no complications). Let $\{ u_{i,t_0} ; i = 1, 
\dots N \}$ be an orthonormal basis for
$\clh^+(t_0)$ at some time $t_0$, and let $\{ v_{i,t_0} ; i = 1 \dots N \}$ be an
orthonormal basis for $\clh^-(t_0)$. The vacuum at time $t_0$ can be written as:
\begin{equation} |{\rm vac}_{t_0} \ra = v_{1,t_0} \wedge \dots \wedge v_{N,t_0} 
\end{equation}
	This state is independent of the choice of basis for $\clh^-(t_0)$ 
(up to a phase factor) because of the complete antisymmetry of the Slater determinant.  If $\hat{H}_1(t)$  depends
	on time then so does the space $\clh^-(t)$, and the
	vacuum $ | {\rm vac}_t \ra$ will differ at different times. If 
$\{ u_{i,t_1} ; i = 1, \dots N \}$ and $\{ v_{i,t_1} ;
i = 1, \dots N \}$ are orthonormal bases for $\clh^+(t_1)$ and
$\clh^-(t_1)$ respectively for some time $t_1 > t_0$, then we may write 
the vacuum at time $t_1$ as:
\begin{equation} |{\rm vac}_{t_1} \ra = v_{1,t_1} \wedge \dots \wedge v_{N,t_1} 
\end{equation} 

	The evolved state 
$| {\rm vac}_{t_0}(t_1) \ra$, obtained by evolving $|{\rm vac}_{t_0} \ra$ from time $t_0$ to
	time $t_1$ is simply:

\begin{equation} |{\rm vac}_{t_0}(t_1) \ra = v_{1,t_0}(t_1) \wedge \dots 
\wedge v_{N,t_0}(t_1) \end{equation}
 where $v_{i,t_0}(t_1)$ denotes the state obtained from
$v_{i,t_0}$ by evolution to time $t_1$, and will not, in general, be 
contained in $\clh^-(t_1)$. We can now calculate quantities such as the 
probability that $| {\rm vac}_{t_0}(t_1) \ra$ will still be in the vacuum state:
\begin{align} \clp_{{\rm vac} \rightarrow {\rm vac}} & = |\la {\rm vac}_{t_1} | 
{\rm vac}_{t_0}(t_1) \ra |^2 \label{eq:vacprob} \\
 \mbox{ where } \la {\rm vac}_{t_1} | {\rm vac}_{t_0}(t_1) \ra & = \det [ 
\la v_{i,t_1} | v_{j,t_0}(t_1) \ra ] \hs{.5} \mbox{from} (\ref{eq:inprod}) \label{eq:vacvac} \end{align}

	In the case of Dirac theory in an electromagnetic background, 
 the time dependence of 
$\hat{H}(t)$ arises entirely from the time dependence of the background $A^{\mu}(\bx,t)$, 
so that the state $| {\rm vac}_{t_0} \ra$ represents the `vacuum in the presence of a 
background $A^{\mu}(\bx,t_0)$', while $| {\rm vac}_{t_1} \ra$ represents the `vacuum in 
the presence of a 
background $A^{\mu}(\bx,t_1)$'. The state
	$ |{\rm vac}_{t_0}(t_1) \ra$, sometimes called the {\it evolved vacuum}, is
	not actually a vacuum state.

\subsection{Particle States Built On The Vacuum}

Let $u_{t_0} \in \clh^+(t_0)$, and let 
$v_{t_0} \in \clh^-(t_0)$.\newline

Then a {\it one-electron state at time $t_0$} 
is of the form: \hs{1} $u_{t_0} \wedge | {\rm vac}_{t_0} \ra$ \newline

while a {\it one-positron state at time $t_0$} is of the form: \hs{1}
$i_{v_{t_0}} | {\rm vac}_{t_0} \ra$ \newline
As expected, electrons are
represented by the presence of positive energy degrees of freedom,
and positrons are represented by the absence of negative energy
degrees of freedom (note that $v_{t_0} \wedge | {\rm vac}_{t_0} \ra = 0 = i_{u_{t_0}} 
| {\rm vac}_{t_0} \ra$ for all $u_{t_0} \in \clh^+(t_0)$ and 
$v_{t_0} \in \clh^-(t_0)$). States having higher numbers of particles 
are constructed in the obvious way. These states are related to the conventional 
constructions by defining creation and
annihilation operators as: 
\begin{align} a(u_{t_0}) = i_{u_{t_0}}  & \hs{3} 
a^{\dagger}(u_{t_0}) =
u_{t_0} \wedge \label{eq:SSB11}\\ 
b(v_{t_0}) = v_{t_0} \wedge  & \hs{3} 
b^{\dagger}(v_{t_0}) = i_{v_{t_0}} \label{eq:SSB12}\end{align} 
	It is routine to verify that $a^{\dagger}(u_{t_0})$ is 
indeed the Hermitian conjugate of $a(u_{t_0})$, and similarly 
for $b^{\dagger}(v_{t_0})$ and $b(v_{t_0})$. The Canonical 
Anticommutation Relations (CARs) follow directly from (\ref{eq:CARprim}), while 
equations such as $$ [ \hat{P}_{\mu}
	, a^{\dagger}_{\lambda}(\bp) ] = p_{\mu}
	a^{\dagger}_{\lambda}(\bp) \hs{1} [ \hat{Q} ,
	a^{\dagger}_{\lambda}(\bp) ] = a^{\dagger}_{\lambda}(\bp)$$
are all contained in Property 2 of Section 2.2.

	It is useful to introduce an orthonormal basis for $\clf_{\wedge}(\clh)$ 
in `standard form' at some time $t_0$. For this purpose, 
 let $\{ u_{i,t_0} ; i = 1, \dots N \}$, $\{ v_{i,t_0} ; i = 1, 
\dots N \}$  be orthonormal bases for $\clh^+(t_0)$ and $\clh^-(t_0)$ respectively, 
as in Section 3.1. Introduce 
the symbol $| \mbox{$\binom{i_1 i_2 \dots i_m}{j_1 j_2 
\dots j_n}$}_{t_0} \ra$ to denote an `in' state, of $m$ particles 
(in states $u_{i_1} \dots u_{i_m}$ with $i_1 < i_2 < \dots i_m$ by 
convention), and $n$ antiparticles (corresponding to 
the absence of states $v_{j_1} \dots v_{j_n}$), prepared at time $t_0$. This 
state is given by:

\begin{align} | \mbox{$\binom{i_1 i_2 \dots i_m}{j_1 j_2 \dots j_n}$}_{t_0} 
\ra & \equiv (-)^J u_{i_1,t_0} \wedge \dots
 u_{i_m,t_0} \wedge v_{1,t_0} \wedge \dots
 \check{v}_{j_1,t_0} \dots \wedge \check{v}_{j_n,t_0} \dots \wedge
 v_{N,t_0} \label{eq:instate1} \\
 & = a^{\dagger}_{i_1,t_0} \dots a^{\dagger}_{i_m,t_0} b^{\dagger}_{j_n,t_0} 
\dots b^{\dagger}_{j_1,t_0} | {\rm vac}_{t_0} \ra \label{eq:instate2} \end{align}
	where the check over $v_{j,t_0}$ signifies that this degree of freedom
 is missing from the state, and $J = \frac{n}{2}(n+1) + \sum_{k=1}^n j_k$ is included as a sign convention. This 
state evolves into
\begin{align} | \mbox{$\binom{i_1 i_2 \dots i_m}{j_1 j_2 \dots j_n}$}_{t_0}(t) \ra & = 
(-)^J u_{i_1,t_0}(t) \wedge \dots \wedge u_{i_m,t_0}(t) 
\wedge v_{1,t_0}(t) \wedge \dots \notag \\
& \hs{1} \dots \check{v}_{j_1,t_0}(t) \dots \wedge \check{v}_{j_n,t_0}(t) 
\dots \wedge v_{N,t_0}(t) \label{eq:inevol} \end{align}
	which we can write loosely as $a^{\dagger}_{i_1,t_0}(t)
	\dots a^{\dagger}_{i_m,t_0}(t) b^{\dagger}_{j_n,t_0}(t) \dots
	b^{\dagger}_{j_1,t_0}(t) |{\rm vac}_{t_0}(t) \ra$ where
	$a^{\dagger}_{i,t_0}(t)$ is shorthand for $u_{i,t_0}(t)
	\wedge$ and $b^{\dagger}_{j,t_0}(t)$ is shorthand for
	$i_{v_{j,t_0}(t)}$. However,
	$a^{\dagger}_{i,t_0}(t)$ does {\it not} represent the creation
	of a particle at time $t$, since $u_{i,t_0}(t)$ is not
	in $\clh^+(t)$. Similarly, $b^{\dagger}_{j,t_0}(t)$ does not
	represent the creation of an antiparticle.

We now define the {\it Time-dependent Bogoliubov Coefficients} by: 
\bea \alpha_{i j}(t,t_0) = \la u_{i,t} | u_{j,t_0}(t) \ra & \hs{1}
	\gamma_{i j}(t,t_0) =  \la v_{i,t} | u_{j,t_0}(t) \ra
	\label{eq:SSB39}\\ \beta_{i j}(t,t_0) = \la u_{i,t} |
	v_{j,t_0}(t) \ra & \hs{1} \epsilon_{i j}(t,t_0) = \la v_{i,t} |
	v_{j,t_0}(t) \ra \label{eq:SSB40}\eea
Then the {\it Bogoliubov conditions}
\begin{equation} \hs{-.3} \left[ \begin{array}{cc} 
 \balpha^{\dagger} \balpha + \bgamma^{\dagger} \bgamma &
 \balpha^{\dagger} \bbeta + \bgamma^{\dagger} \bepsilon\\
 \bbeta^{\dagger} \balpha + \bepsilon^{\dagger} \bgamma &
 \bbeta^{\dagger} \bbeta + \bepsilon^{\dagger} \bepsilon \end{array}
 \right] =   \left[ \begin{array}{cc}  \bI & 0 \\ 0 & \bI \end{array}
 \right] =  \left[ \begin{array}{cc}  \balpha \balpha^{\dagger} +
 \bbeta \bbeta^{\dagger} & \balpha \bgamma^{\dagger} + \bbeta
 \bepsilon^{\dagger} \\ \bgamma \balpha^{\dagger} + \bepsilon
 \bbeta^{\dagger} & \bgamma \bgamma^{\dagger} + \bepsilon
 \bepsilon^{\dagger} \end{array} \right]
\label{eq:SSB43}\end{equation}
	follow from unitarity of the `first quantized' evolution 
matrix 
$$ \bS_1(t_1,t_0) = \left[ \begin{array}{cc}  \balpha(t_1,t_0) & \bbeta(t_1,t_0) \\
 \bgamma(t_1,t_0) & \bepsilon(t_1,t_0) \end{array} \right]$$ Note that our labelling of the Bogoliubov coefficients 
differs from that used in Manogue's \cite{Man} asymptotic treatment (Manogue's are the
complex conjugates of those used here). From (\ref{eq:inprod}), (\ref{eq:instate1}) and (\ref{eq:inevol}) 
we have immediately:
\begin{align} & \la \mbox{$\binom{i'_1 i'_2 \dots 
i'_{m'}}{j'_1 j'_2 \cdots j'_{n'}}$}_{t_1} | \mbox{$\binom{i_1 i_2 \dots 
i_m}{j_1 j_2 \dots j_n}$}_{t_0}(t_1) \ra \notag \\
 & = (-)^{J-J'} \det \hs{-.1} \left[ \hs{-.1} \begin{array}{cc}  \left[ \begin{array}{ccc}
  \alpha_{i'_1 i_1} & \dots & \alpha_{i'_1 i_m} \\ \vdots & & \vdots
  \\ \alpha_{i'_{m'} i_1}  & \cdots & \alpha_{i'_{m'} i_m} \end{array}
  \right] &  \left[ \begin{array}{ccc}  \beta_{i'_1 1} & \binom{j_1
  \dots j_n}{missing} & \beta_{i'_1 N} \\ \vdots & & \vdots \\
  \beta_{i'_{m'} 1}  & \binom{j_1 \dots j_n}{missing} & \beta_{i'_{m'}
  N} \end{array} \right] \\ 
\hs{-.1} \left[ \hs{-.1} \begin{array}{ccc}  \gamma_{1 i_1}
  & \hs{-.1} \cdots \hs{-.1} & \gamma_{1 i_m} \\ \binom{j'_1 \dots j'_{n'}}{missing} & &
  \binom{j'_1 \dots j'_{n'}}{missing} \\  \gamma_{N i_1}  & \hs{-.1} \cdots \hs{-.1} &
  \gamma_{N i_m} \end{array} \hs{-.1} \right] \hs{-.2} &  \hs{-.2} \left[ \hs{-.1} \begin{array}{ccc}
  \epsilon_{1 1} & \hs{-.1} \binom{j_1 \dots j_n}{missing} \hs{-.1} & \epsilon_{1 N} \\
  \binom{j'_1 \dots j'_{n'}}{missing} & & \binom{j'_1 \dots
  j'_{n'}}{missing} \\ \epsilon_{N 1} & \hs{-.1} \binom{j_1 \dots j_n}{missing} \hs{-.1}
  & \epsilon_{N N} \end{array} \hs{-.1} \right] \hs{-.1} \end{array} \hs{-.1} \right] \hs{-.1}
  \label{eq:SSBSmat}\end{align} if $m-n = m'-n'$ (charge
  conservation), and zero otherwise. This is a completely
  general formula for an arbitrary S-Matrix element, and it is 
remarkable that its derivation is so easy. This is in stark contrast to 
conventional formulations (see Section 5). The reason is the
concrete representation of states as given 
by (\ref{eq:instate1}), and the simple evolution equation which allows 
us to deduce (\ref{eq:inevol}). In most approaches to QFT 
in a classical background \cite{BD,Full,DeWitt} the
	states are defined only implicitly, by equations like (\ref{eq:instate2}), 
and the creation/annihilation operators are defined implicitly, 
by the CAR's. The derivation of S-Matrix elements must then proceed  by a much more round-about method.

	Some other textbook approaches do present a concrete 
representation of state space \cite{Thal,Wa}, and also use the 
antisymmetric tensor algebra $\clf_{\wedge}(\clh')$ over 
some `first quantised' state space $\clh'$. However, they do not simply 
take $\clh'$ to be the space of Dirac states, but rather require that 
it represent the `set of all one-particle states'. Then $\wedge^n \clh'$
	represents the set of all $n$-particle states, while $\wedge^0 \clh'$ 
represents the physical vacuum. Solutions of 
	the Dirac equation must then be modified in order to construct these 
	1-particle states. For example, Thaller (\cite{Thal}, page 275) 
	uses the set of `positive energy solutions and charge
	conjugates of negative energy solutions' $\clh' \equiv \clh^+ + 
\hat{C} \clh^-$, but treats only those cases where this choice can be
	made independently of time, while Wald (\cite{Wa}, page 103) uses
	a construction that no longer makes reference
	to positive or negative energy states. Such modifications of $\clh$ 
destroy the simplicity of the evolution equation (\ref{eq:evol}) and require  
that grade violating terms be included to represent pair creation. This is why, 
even with a concrete representation of state space, such approaches still find 
it easier to describe those states in terms of creation/annihilation operators 
and a field operator rather than referring to the states directly, and why 
(\ref{eq:SSBSmat}) is derived using methods like those in Section 5. Also, the requirement  
that $\clh'$ be time independent either restricts the 
applications of the theory or implies that $\clh'$ can no longer 
be interpreted as representing 1-particle states, removing any  
motivation for such a choice of $\clh'$.\newline

	Before we examine expectation values and vacuum subtraction, 
it is convenient to factor out $\la {\rm vac}_{t_1} | {\rm vac}_{t_0}(t_1) \ra = \det(\bepsilon(t_1,t_0))$ 
from some of the S-Matrix elements presented in (\ref{eq:SSBSmat}). Consider for 
example $\la \mbox{$\binom{i'_1 i'_2 \dots i'_{n'}}{j'_1 j'_2 \dots j'_{n'}}$}_{t_1} | {\rm vac}_{t_0}(t_1) \ra$. This can be written as:
\vs{-.5}

\begin{equation} \la \mbox{$\binom{i'_1 i'_2 \dots i'_{n'}}{j'_1 j'_2 \dots j'_{n'}}$}_{t_1} | {\rm vac}_{t_0}(t_1) \ra = (-)^{J'} \det \left[ \begin{array}{c}
 \ora{\beta}_{i'_1} \\ \vdots \\ \ora{\beta}_{i'_{n'}} \\
 \ora{\epsilon}_1 \\ \binom{j'_1 \dots j'_{n'}}{missing} \\
 \ora{\epsilon}_N \end{array} \right] \notag \end{equation} 
\vs{-.2}

where
 $\ora{\beta}_{i}$ represents the $i^{th}$ row of $\bbeta$, and
 similarly for $\ora{\epsilon}_i$. By multiplying this on the right by
 the matrix $\left[ \begin{array}{cccccc}\ora{\epsilon}^{-1}_{j'_1} &
 \cdots & \ora{\epsilon}^{-1}_{j'_{n'}} & \ora{\epsilon}^{-1}_{1} &
 \binom{j'_1 \dots j'_{n'}}{missing} & \ora{\epsilon}^{-1}_{N}
 \end{array} \right]$ (having determinant 
 $\frac{(-1)^{\sum_1^{n'} (j'_k - 1)}}{\det(\bepsilon)}$) we have
\begin{align} \la \mbox{$\binom{i'_1 i'_2 \dots i'_{m'}}{j'_1
 j'_2 \dots j'_{n'}}$}_{t_1} | {\rm vac}_{t_0}(t_1) \ra
 & = (-)^{\frac{n'}{2}(n'-1)} \det(\bepsilon) \det \left
 [ \begin{array}{ccc}  V_{i'_1 j'_1} & \cdots & V_{i'_1 j'_{n'}} \\
 \vdots & & \vdots \\ V_{i'_{n'} j'_1} & \cdots & V_{i'_{n'} j'_{n'}}
 \end{array} \right] \notag \\ 
& = (-)^{\frac{n'}{2}(n'-1)}
 \det(\bepsilon) \det(\bV_{out})
 \label{eq:SSBScase1} \end{align} where $\bV \equiv \bbeta
 \bepsilon^{-1}$ and $\bV_{out}$ is constructed from $\bV$ using only
 those matrix entries relevant to the desired out state. Similarly:

\begin{align} \la {\rm vac}_{t_1} |
 \mbox{$\binom{i_1 i_2 \dots i_m}{j_1 j_2 \dots j_n}$}_{t_0}(t_1) \ra
  & = (-)^{\frac{n}{2}(n-1)}\det(\bepsilon) \det \left[
 \begin{array}{ccc}  \Lambda_{j_1 i_1} & \cdots & \Lambda_{j_1 i_{n}}
 \\ \vdots & & \vdots \\ \Lambda_{j_{n} i_1} & \cdots & \Lambda_{j_{n}
 i_{n}} \end{array} \right]  \label{eq:SSBScase2} \end{align}
 where $\bLambda \equiv \bepsilon^{-1} \bgamma$.

The general case is more arduous, but routine. Rather than present it here we shall  
consider, as an example, the S-Matrix element between 
two 1-electron states, $\la \mbox{$\binom{i'}{ }$}_{t_1} 
| \mbox{$\binom{i}{ }$}_{t_0}(t_1) \ra  = 
\la {\rm vac}_{t_1} |a^{\dagger}_{i',t_1} a^{\dagger}_{i,t_0}(t_1)|{\rm vac}_{t_0}(t_1) \ra$. We 
first expand $u_{i,t_1}$ in terms of $\{ u_{j,t_0}(t_1), v_{j,t_0}(t_1) \}$, as 
$u_{i,t_1} = \sum_j \{ \alpha^*_{i j}(t_1,t_0) u_{j,t_0}(t_1) +  
\beta^*_{i j}(t_1,t_0) v_{j,t_0}(t_1) \}$. From the definition (\ref{eq:SSB11}), 
this immediately yields the familiar equation:
\begin{equation} a_{i,t_1} = \sum_j \{ \alpha_{i j}(t_1,t_0) a_{j,t_0}(t_1) + 
\beta_{i j}(t_1,t_0) b^{\dagger}_{j,t_0}(t_1) \} \label{eq:SSB48new} \end{equation}
	From this, we can write:

\begin{align} \la \mbox{$\binom{i'}{ }$}_{t_1} 
| \mbox{$\binom{i}{ }$}_{t_0}(t_1) \ra 
 & = \la {\rm vac}_{t_1} | \alpha_{i' i} + \sum_j 
\beta_{i' j} b^{\dagger}_{j,t_0}(t_1) a^{\dagger}_{i,t_0}(t_1)  |{\rm vac}_{t_0}(t_1) \ra
\label{eq:gh}\\  &  = \det(\bepsilon) (\balpha - \bbeta
 \bepsilon^{-1} \bgamma)_{i' i} = \overline{\alpha^{-1}_{i i'}}
 \det(\bepsilon) \label{eq:part/part}\end{align} where we have used $\balpha - \bbeta \bepsilon^{-1} \bgamma = (\balpha^{\dagger})^{-1}$,
 which follows from the Bogoliubov conditions. The two terms in 
(\ref{eq:gh}) represent distinct contributions to the `evolved 
1-particle state' $u_{i,t_0}(t_1) \wedge | {\rm vac}_{t_0}(t_1) \ra$. The first 
contribution corresponds to the positive energy component of $u_{1,t_0}(t_1)$ acting on 
the vacuum component of
$| {\rm vac}_{t_0}(t_1) \ra$, while the second contribution comes from the 
negative energy component of $u_{1,t_0}(t_1)$ destroying the 
antiparticle of one of the two-particle components of $| {\rm vac}_{t_0}(t_1) 
\ra$. Similarly, the S-matrix element between
 two antiparticle states is:
\begin{equation} \la \mbox{$\binom{ }{j'}$}_{t_1} | \mbox{$\binom{ }{j}$}_{t_0}(t_1) \ra =  (\epsilon^{-1})_{j j'} \det(\bepsilon) \label{eq:anti/anti}
\end{equation}

	We can use the Bogoliubov conditions to deduce the relations
\begin{align} \bLambda^{\dagger} = - \balpha^{-1} \bbeta \hs{2} & \bV^{\dagger} = - 
\bgamma \balpha^{-1} \label{eq:Vdagger1}\\
	\bepsilon^{\dagger} \bepsilon = (1 + \bLambda
	\bLambda^{\dagger})^{-1} \hs{2} & \balpha \balpha^{\dagger} =
	(1 + \bV \bV^{\dagger})^{-1} \label{eq:Vdagger2}\end{align}
	Since $\balpha^{-1}$ and $\bepsilon^{-1}$ respectively represent the
	relative particle/particle and antiparticle/antiparticle
	transition amplitudes, it follows that (\ref{eq:Vdagger2}),
	along with the relations $\bV^{\dagger} \balpha = - \bepsilon \bLambda$ and
	$\balpha \bLambda^{\dagger} = - \bV \bepsilon$ (these follow
	from (\ref{eq:Vdagger1}) and the definitions of
	$\bV$, $\bLambda$) constitute the fermionic version of
	the relativistic generalisation
	of the Optical Theorem, that has been given by DeWitt \cite{DeWitt,DeWitt2}.

\subsection{Vacuum Subtraction and Physical Operators} 

	Consider the expectation value, in the physical vacuum at 
time $t_0$, of the Hermitian extension of an operator 
$\hat{A}_1(t_0):\clh \rightarrow \clh$. From (\ref{eq:prop3a}) we have:
\begin{equation} \la {\rm vac}_{t_0} |\hat{A}_H(t_0)| {\rm vac}_{t_0} \ra = 
\sum_i \la v_{i,t_0} |\hat{A}_1(t_0)| v_{i,t_0} \ra \notag \end{equation}
where $\{ v_{i,t_0} \}$  constitute an orthonormal basis for $\clh^-(t_0)$. This is not 
zero, nor even finite in general. At this point we need to introduce 
a scheme analogous to normal ordering. Accordingly, given an operator 
$\hat{A}_1(t):\clh \rightarrow \clh$  corresponding to an observable which is 
zero in the vacuum we define the {\it physical extension} 
$\hat{A}_{{\rm phys}}(t): \clf_{\wedge}(\clh) \rightarrow 
\clf_{\wedge}(\clh)$ of $\hat{A}_1(t)$ by:

 \begin{equation} \hat{A}_{{\rm phys}}(t) = \hat{A}_H(t) - \la {\rm vac}_{t} | \hat{A}_H(t) | {\rm vac}_{t} \ra \hat{1} \label{eq:vacsub1} \end{equation}
	So  $ \la {\rm vac}_t | \hat{A}_{{\rm phys}}(t) | {\rm vac}_t \ra$ is now zero by construction. 

In Section 5.2 we show that 
this vacuum subtraction is equivalent to normal ordering with respect to the particle 
interpretation at the time of measurement. This choice is also made in  
previous `Hamiltonian diagonalisation' procedures, and uniquely  
guarantees the positive definiteness of $\hat{H}_{{\rm phys}}(t)$ while maintaining 
$\la {\rm vac}_t | \hat{H}_{{\rm phys}}(t) | {\rm vac}_t \ra = 0$. Unfortunately, experience has 
shown that vacuum subtraction does not on its own always return a finite  
expectation value of $\hat{H}_{{\rm phys}}(t)$ in the {\it evolved vacuum}, $\la {\rm vac}_{t_0}(t) 
| \hat{H}_{{\rm phys}}(t) | {\rm vac}_{t_0}(t) \ra$. Further renormalisation is often
required. Techniques for the renormalisation of 
quantities `after 
vacuum subtraction' are presented in, for example, Grib et al.\cite{GMM}, 
while many other techniques are presented in \cite{BD} or \cite{Full}.

	We now treat two important examples of `physical
	extension'.

\begin{enumerate}

\item Consider the unit operator on $\clh$. Its physical
extension clearly satisfies:

\begin{align} 
\hat{1}_{{\rm phys}} | \mbox{$\binom{i_1 i_2 \dots i_m}{j_1 j_2 \dots j_n}$}_{t_0}(t) \ra 
& = \{ {\rm grade}(| \mbox{$\binom{i_1 i_2 \dots i_m}{j_1 j_2 \dots j_n}$}_{t_0}(t) \ra) 
- {\rm grade}(| {\rm vac}_t \ra) \} | \mbox{$\binom{i_1 i_2 \dots i_m}{j_1 j_2 \dots j_n}$}_{t_0}(t) \ra \notag \\
 & = (m - n) | \mbox{$\binom{i_1 i_2 \dots i_m}{j_1 j_2 \dots j_n}$}_{t_0}(t) \ra \label{eq:SSB19} \notag \end{align}

 at all times $t$ (including $t=t_0$). 
This clearly
 represents charge; $\hat{Q} = e \hat{1}_{{\rm phys}}$. Charge conservation 
therefore follows directly from the fact that evolution is
 grade preserving. That the physical extension of the unit operator should 
represent charge is clearly
 appropriate, since the norm $\la \psi | \hat{1}_1 | \psi \ra$ of a
 state $\psi \in \clh$ is well known to be the conserved `charge' of
 the Dirac Lagrangian conjugate to changes in phase.

\item The number operator is the most important example of an operator 
depending explicitly on the split of $\clh$ into $\clh^+(t_0)$ and
$\clh^-(t_0)$. It is the physical extension of the operator $\hat{N}_1(t_0) :\clh \rightarrow
\clh$ defined by $\hat{N}_1(t_0) = \hat{P}^+(t_0) -
\hat{P}^-(t_0)$. Clearly $\hat{N}_1(t_0)$ commutes with $\hat{H}_1(t_0)$, but does not
in general commute with time evolution (since it does not commute with 
$\hat{H}_1(t)$ for $t \neq t_0$). $\hat{N}_{{\rm phys}}(t_0)$ inherits both of 
these properties. Therefore the number operator $\hat{N}_{{\rm phys}}(t_0)$
represents a well-defined physical observable, but is not
conserved. When acting on states in standard form, it gives:

\begin{equation} \hat{N}_{{\rm phys}}(t_0) | \mbox{$\binom{i_1 i_2 \dots i_m}{j_1 j_2 \dots j_n}$}_{t_0} \ra = (m + n) | \mbox{$\binom{i_1 i_2 \dots i_m}{j_1 j_2 \dots j_n}$}_{t_0} \ra 
 \label{eq:SSB22}\end{equation}
	so that it is positive definite and has integer eigenvalues. We can 
expand $\hat{N}_{{\rm phys}}(t_0)$ in modes as: 
$$\hat{N}_{{\rm phys}}(t_0) = \sum_{k} (\hat{N}^+_{k,t_0} + \hat{N}^-_{k,t_0})$$ 
where $\{ u_{k,t_0} \}$ and $\{ v_{k,t_0} \}$ are orthonormal 
bases of $\clh^{\pm}(t_0)$, \linebreak $\hat{N}^+_{k,t_0} = u_{k,t_0} \wedge (i_{u_{k,t_0}}$ 
is the physical extension 
of $| u_{k,t_0} \ra \la u_{k,t_0} |$ and  \linebreak $\hat{N}^-_{k,t_0} = i_{v_{k,t_0}} (v_{k,t_0} \wedge \hs{.1} $
 is the physical extension of $ - | v_{k,t_0} \ra \la v_{k,t_0} |$. Similarly: 
$$\hat{Q}_{{\rm phys}} = e \sum_{k} (\hat{N}^+_{k,\tau} - \hat{N}^-_{k,\tau})$$.

\end{enumerate}

\subsection{Expectation Values}

Consider the expectation values of an operator $\hat{A}_{{\rm phys}}(t_1)$ which 
is the physical extension of some operator
 $\hat{A}_1(t_1)$. Consider first, for simplicity, its expectation 
value in the so-called `evolved vacuum' $|{\rm vac}_{t_0}(t_1) \ra$. From 
(\ref{eq:prop3a}) and (\ref{eq:vacsub1}) we have
\begin{equation} \la {\rm vac}_{t_0}(t_1) | \hat{A}_{{\rm phys}}(t_1) | {\rm vac}_{t_0}(t_1) \ra = 
\sum_{i=1}^{N} \la v_{i,t_0}(t_1) | \hat{A}_1(t_1) | v_{i,t_0}(t_1) \ra - \sum_{i=1}^{N} 
\la v_{i,t_1} | \hat{A}_1(t_1) | v_{i,t_1} \ra  \label{eq:vacsub} \end{equation}
This can be expressed in terms of Bogoliubov coefficients by inserting factors of $\hat{1} = 
\sum_{i} \{ |u_{i,t_1} \ra \la u_{i,t_1} | + | v_{i,t_1} \ra \la v_{i,t_1} | \}$  
on either side of $\hat{A}_1(t_1)$ in the first term and rearranging, to give
\begin{align} \la {\rm vac}_{t_0}(t_1) & | \hat{A}_{{\rm phys}}(t_1) | {\rm vac}_{t_0}(t_1) \ra \notag \\
 & \hs{-.5} = {\rm Trace}( \bbeta^{\dagger} \bA^{++} \bbeta + \bbeta^{\dagger} \bA^{+-} \bepsilon + 
\bepsilon^{\dagger} \bA^{-+} \bbeta + \bepsilon^{\dagger} \bA^{--} \bepsilon ) -
{\rm Trace}(\bA^{--}) \notag \\
 & \hs{-.5} = {\rm Trace}(\bbeta \bbeta^{\dagger} \bA^{++} - \bgamma
	\bgamma^{\dagger} \bA^{--} + \bepsilon \bbeta^{\dagger}
	\bA^{+-} + \bbeta \bepsilon^{\dagger} \bA^{-+})
	\label{eq:SSB40.1} \end{align} 
where we have defined:
\begin{align} \bA^{++}_{j k} \equiv \la u_{j,t_1} | \hat{A}_1(t_1) | u_{k,t_1} \ra 
 \hs{1} & \bA^{--}_{j k} \equiv \la v_{j,t_1} | \hat{A}_1(t_1) | v_{k,t_1} \ra \notag \\ 
\bA^{+-}_{j k} \equiv \la u_{j,t_1} | \hat{A}_1(t_1) |
  v_{k,t_1} \ra \hs{.2} \mbox{ and } & \bA^{-+}_{j k} \equiv \la
  v_{j,t_1} | \hat{A}_1(t_1) | u_{k,t_1} \ra \label{eq:SSB40.2} \\
 & \hs{.8} = \overline{\bA^{+-}_{k j}} \mbox{ if } \hat{A}_1 \mbox{ is Hermitian
  } \notag \end{align} 
	The
  subtraction $(\bepsilon \bepsilon^{\dagger})_{k j} - \delta_{k j} =
  - (\bgamma \bgamma^{\dagger})_{k j}$, which is used to pass to (\ref{eq:SSB40.1})
  relies on the fact that we are vacuum
  subtracting with respect to the vacuum {\it at the time of 
measurement}.

	As a simple example of (\ref{eq:SSB40.1}) we can 
calculate the number of particles in the evolved 
vacuum by using $\hat{N}_1(t_1) = \hat{P}^+(t_1)
  - \hat{P}^-(t_1)$, so that $\bN^{++}_{j k} = \delta_{j k} = -
  \bN^{--}_{j k}$ and $\bN^{+-}_{j k} = 0 = \bN^{-+}_{j k}$. This gives:
\begin{align} N_{{\rm vac},t_0}(t_1) & \equiv \la {\rm vac}_{t_0}(t_1)| \hat{N}_{{\rm phys}}(t_1) | {\rm vac}_{t_0}(t_1) \ra = {\rm Trace}( \bbeta \bbeta^{\dagger} + \bgamma
	\bgamma^{\dagger}) \notag \\
& = 2 {\rm Trace}( \bbeta \bbeta^{\dagger}) \end{align}
 where we have used the Bogoliubov conditions in the last stage. 
Conservation of charge follows by considering
  $\hat{A}_1 = \hat{1}$. Also, we can now write equation (\ref{eq:vacprob}) as:
\begin{align} \clp_{|{\rm vac}_{t_0}\ra \rightarrow |{\rm vac}_{t_1}\ra} & = |\det( \bepsilon(t_1,
t_0))|^2 = \det(\bepsilon(t_1,t_0)\bepsilon^{\dagger}(t,t_0)) \label{eq:SSB57} \\
 & = \exp({\rm Trace}(\log(1 - \bgamma \bgamma^{\dagger})) = \exp(-
\sum_{n=1}^{\infty} \frac{1}{n} {\rm Trace}((\bgamma \bgamma^{\dagger})^n))
\label{eq:SSB59}\end{align} In an electromagnetic
background $\bgamma$ is first order in the coupling
constant $e$, so that $\bgamma \bgamma^{\dagger}$ is second
order. Hence (\ref{eq:SSB59}) allows us to write: \begin{equation}
\clp_{|{\rm vac}_{t_0}\ra \rightarrow |{\rm vac}_{t_1}\ra} = \exp( - \half N_{{\rm vac},t_0}(t_1) ) 
+ O(e^4) \label{eq:SSB66}\end{equation}  which relates the 
probability of vacuum decay to the expected pair creation. This relation is used in
Itzykson and Zuber \cite{IZ} to 
find $N_{{\rm vac},-\infty}(\infty)$ (to this order in $e$) without using
Bogoliubuv coefficients. However Itzykson and Zuber \cite{IZ} justify the result on physical grounds, without formal proof. Perturbative 
calculations now involve simply the expansion of the `first
quantized' solutions  $u_{i,t_0}(t), v_{i,t_0}(t) $ in powers of the coupling
constant $e$, so as to generate expansions of the Bogoliubov 
coefficients. This is considered in detail in \cite{mythesis}.

	The derivation of $\la F_{t_0}(t_1) |
  \hat{A}_{{\rm phys}}(t_1) | F_{t_0}(t_1) \ra$ for an arbitrary state $| F_{t_0}(t_1) \ra$
  is identical to the derivation of (\ref{eq:SSB40.1}), and gives:

\begin{align} & \la \mbox{$\binom{i_1 i_2 \dots i_m}{j_1 j_2 \dots j_n}$}_{t_0}(t_1)|
 \hat{A}_{{\rm phys}}(t_1) | \mbox{$\binom{i_1 i_2 \dots i_m}{j_1 j_2 \dots j_n}$}_{t_0}(t_1)
 \ra \notag \\
 & \hs{1} = \sum_{k=1}^{m} \la u_{i_k,t_0}(t_1) | \hat{A}_1(t_1) |
 u_{i_k,t_0}(t_1) \ra - \sum_{k=1}^{n} \la v_{j_k,t_0}(t_1) | \hat{A}_1(t_1)
 | v_{j_k,t_0}(t_1) \ra \notag \\
 & \hs{2} + \la {\rm vac}_{t_0}(t_1) | \hat{A}_{{\rm phys}}(t_1)
 | {\rm vac}_{t_0}(t_1) \ra \label{eq:SSB40.12}\\ 
& \hs{1} = \sum_{k=1}^{m}
 (\balpha^{\dagger} \bA^{++} \balpha + \balpha^{\dagger} \bA^{+-}
 \bgamma + \bgamma^{\dagger} \bA^{-+} \balpha + \bgamma^{\dagger}
 \bA^{--} \bgamma )_{i_k i_k} \notag \\ 
& \hs{2} - \sum_{k=1}^{n}
 (\bbeta^{\dagger} \bA^{++} \bbeta + \bbeta^{\dagger} \bA^{+-}
 \bepsilon + \bepsilon^{\dagger} \bA^{-+} \bbeta + \bepsilon^{\dagger}
 \bA^{--} \bepsilon )_{j_k j_k} \notag \\
& \hs{2} + \la {\rm vac}_{t_0}(t_1) | \hat{A}_{{\rm phys}}(t_1)
 | {\rm vac}_{t_0}(t_1) \ra \label{eq:SSB40.15} \end{align}

\subsubsection*{Anomalies and Fluctuations}

	Now let $\hat{A}_1(t)$
represent a quantity that is conserved at the level of the Dirac
equation. In this case $\la v_{i,t_0}(t_1) | \hat{A}_1(t_1) |
v_{i,t_0}(t_1) \ra = \la v_{i,t_0} | \hat{A}_1(t_0) | v_{i,t_0} \ra$, so that 
(\ref{eq:SSB40.1}) becomes:
$$ \la {\rm vac}_{t_0}(t_1) | \hat{A}_{{\rm phys}} | {\rm vac}_{t_0}(t_1) \ra = 
{\rm Trace}(\bA^{--}(t_0) - \bA^{--}(t_1)) $$
	This can be non-zero even when $\hat{A}_{1}$ is independent of
	time, because of the varying particle interpretation. Herein lies an elegant physical description of quantum
	anomalies. Although the expectation value of $\hat{A}_H$ in
	any given state does not change with time, the portion of it
	attributable to the vacuum may change, as the
	definition of the vacuum changes. Since the amount that would be
	measured experimentally is the amount left after subtraction of the vacuum this can change with
	time. As an example, an elegant treatment of the axial anomaly in an
	external electromagnetic background, which appeals to a
	similar physical mechanism as above, is given in
	\cite{AxAn} and in \cite{Jackiw}.

	Fluctuations in expectation 
values can be calculated using (\ref{eq:prop3c}). Consider fluctuations 
of some quantity $\hat{A}(t_1)$ in the `evolved vacuum'. Define 
$$\hat{P}^-_{t_0}(t_1) \equiv \sum_{i} |v_{i,t_0}(t_1) \ra 
\la v_{i,t_0}(t_1) | = \hat{U}_1(t_1,t_0) \hat{P}^-(t_0) 
\hat{U}^{\dagger}_1(t_1,t_0)$$ and similarly for 
$\hat{P}^+_{t_0}(t_1)$. Then we can write:

\begin{align} \la {\rm vac}_{t_0}(t_1) & | (\hat{A}_{{\rm phys}})^2 | {\rm vac}_{t_0}(t_1) \ra
 - \la {\rm vac}_{t_0}(t_1) | \hat{A}_{{\rm phys}} | {\rm vac}_{t_0}(t_1) \ra^2 \\ 
& = \sum_i \la v_{i,t_0}(t_1) | \hat{A}_1(t_1)^2 - \hat{A}_1(t_1) \hat{P}^-_{t_0}(t_1) \hat{A}_1(t_1) |v_{i,t_0}(t_1) \ra \\
 & = {\rm Trace}(\hat{P}^-_{t_0}(t_1) \hat{A}_1(t_1) \hat{P}^+_{t_0}(t_1) \hat{A}_1(t_1)) \end{align}
	where the trace here is 2N-dimensional. It follows that the size 
of the fluctuations in the evolved vacuum are determined by the commutator 
of $\hat{A}_1(t_1)$ and $\hat{P}^-_{t_0}(t_1)$. If we consider 
fluctuations in the physical vacuum at time $t_1$, $| {\rm vac}_{t_1} \ra$ 
(by putting $t_0 = t_1$ above), we observe that any operator commuting with 
the Hamiltonian (and hence with $\hat{P}^-(t_1)$) has zero fluctuations 
in the physical vacuum. This reflects the fact that the 
physical vacuum, which is constructed from the spectrum of the Hamiltonian, 
must be an eigenstate of any operator that commutes with the Hamiltonian.

\section{Arbitrary Observers}

Implicit in the construction presented so far is the assumption that 
the `in-state' is prepared on the spacelike Cauchy surface $t=t_0$ for some $t_0$, 
and that the out state is to be measured on the spacelike Cauchy surface $t=t_1$ 
for some $t_1 > t_0$. We will show in this Section that this 
amounts to assuming that the system is being observed by an 
inertial observer, and described in their rest frame. We also show how 
the particle interpretation can be generalised to an arbitrary observer, 
by using their `radar time', and we will demonstrate that this definition 
depends only on the motion of the observer, not on their choice of coordinates 
or their choice of gauge. The nonlocal nature of this definition will also be discussed.

\subsection{An Observer and The Corresponding Particle Interpretation}

Consider an observer travelling on path $\gamma: x^{\mu} = x^{\mu}(\tau)$ with proper time $\tau$, and define:
\begin{align}
\tau^{+}(x) & \equiv \mbox{ (earliest possible) proper time at which a null geodesic leaving} \notag \\
& \hs{2} \mbox{ point $x$ could intercept $\gamma$. } \notag \\
\tau^{-}(x) & \equiv \mbox{ (latest possible) proper time at which a null geodesic could} \notag \\
& \hs{2} \mbox{ leave $\gamma$, and still reach point $x$. } \notag \\
\tau(x) & \equiv \half (\tau^{+}(x) + \tau^{-}(x)) \hs{1} = \mbox{ `radar time'.} \notag \\
\rho(x) & \equiv \half (\tau^{+}(x) - \tau^{-}(x)) \hs{1} = \mbox{ `radar distance'.} \notag \\
\Sigma_{\tau_0} & \equiv \{x: \tau(x) = \tau_0 \} = \mbox{ observer's `hypersurface 
of simultaneity at time $\tau_0$'. } \notag \end{align}

\begin{figure}[h]
\vspace{-.3cm}
\center{\epsfig{figure=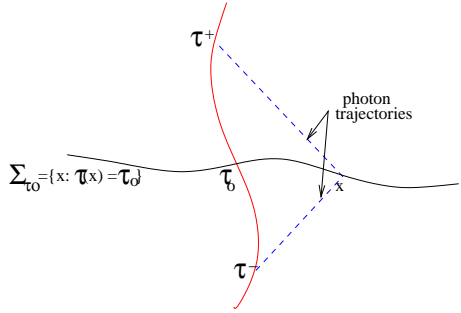, width=6cm}}
\caption{{\footnotesize Schematic of the definition of `radar time' $\tau(x)$.}}
\vspace{-.3cm}
\end{figure}

This is a simple generalisation of the definition made popular by Bondi in his work on 
special relativity and k-calculus \cite{Bondi,Bohm,Dinverno}. It generalises 
immediately to gravitational backgrounds \cite{Me2,mythesis}. It is the 
only possible construction which agrees with proper time on the 
observer's path, and is also invariant under `time-reversal' - that is, 
under reversal of the sign of the observer's proper time.

	We can now define the `time-translation' vector field: 
$$k_{\mu}(x) \equiv \frac{\frac{\partial \tau}{\partial x^{\mu}}}{\eta^{\sigma \nu} 
\frac{\partial \tau}{\partial x^{\sigma}} \frac{\partial \tau}{\partial x^{\nu}}} $$
	This represents the perpendicular distance between neighbouring hypersurfaces of 
simultaneity, since it is normal to these hypersurfaces, and satisfies 
$k^{\mu}(x)\frac{\partial \tau}{\partial x^{\mu}} = 1$. Now use the identity 
$i \hs{.1}|\hs{-.15}k \gamma^{\mu} \grad_{\mu} = i k^{\mu} \grad_{\mu} + 
\sigma^{\mu \nu} k_{\mu} \grad_{\nu}$  (where $\hs{.1}|\hs{-.15}k \equiv k_{\mu} \gamma^{\mu}$ 
and $\sigma^{\mu \nu} \equiv \frac{i}{2} [ \gamma^{\mu} , 
\gamma^{\nu}]$) to write (\ref{eq:diss2.2}) as:
\begin{equation} i k^{\mu} \grad_{\mu} \psi = - \sigma^{\mu \nu} k_{\mu} 
\grad_{\nu} \psi + m \hs{.1}|\hs{-.15}k \psi \label{eq:dissHam2}\end{equation}
	
	From this we can define the `Hamiltonian on $\Sigma_{\tau}$' $\hat{H}_{{\rm nh}}(\tau)$ by:
\begin{equation} \hat{H}_{{\rm nh}}(\tau):\psi \rightarrow - \sigma^{\mu \nu} k_{\mu} 
\grad_{\nu} \psi + m \hs{.1}|\hs{-.15}k \psi \label{eq:Hgen2} \end{equation}

$\hat{H}_{{\rm nh}}(\tau)$ is not in general Hermitian! (hence the 
subscript $nh$ here). At first sight this seems to 
disagree with unitarity on $\clh$. In fact there is no  
inconsistency, because the inner product now depends (via the volume 
element on $\Sigma_t$) explicitly on $\tau$, and because  
(\ref{eq:dissHam2}) is no longer of the form $i \frac{d}{d t} | \psi(t) \ra = \hat{H}_1(t) 
| \psi(t) \ra$ so that the standard equivalence proof of unitary evolution 
 and a Hermitian Hamiltonian no longer applies. 

	To investigate the relation between $\hat{H}_{{\rm nh}}(\tau_0)$ 
and the energy momentum tensor, define:

\begin{equation} H_{\tau_0}(\psi) \equiv \int_{\Sigma_{\tau_0}} T_{\mu \nu}(\psi(x)) k^{\mu} {\rm d} 
\Sigma^{\nu} \label{eq:Hsig} \end{equation}
	By substituting the energy momentum tensor (\ref{eq:enmom}) into this we see that:

\begin{gather} H_{\tau_0}(\psi) = \Re [ \la \psi | \hat{H}_{{\rm nh}}(\tau_0) | \psi 
\ra_{\Sigma_{\tau_0}} ] = \la \psi | \hat{H}_1(\tau_0) | \psi \ra_{\Sigma_{\tau_0}} \label{eq:Hsig2} \\
 \mbox{ where } \hat{H}_1(\tau_0) \equiv \half \{ \hat{H}_{{\rm nh}}(\tau_0) +  
\hat{H}^{\dagger}_{{\rm nh}}(\tau_0) \} \end{gather}
It follows that we can define the projection operators $\hat{P}^{\pm}(\tau_0)$ and the spaces 
$\clh^{\pm}(\tau_0)$ by requiring that $H_{\tau_0}(\hat{P}^+(\tau_0) \psi) \geq H_{\tau_0}(\psi) \geq 
H_{\tau_0}(\hat{P}^-(\tau_0) \psi)$ for all $\psi$, just as in (\ref{eq:Hdef}). Clearly 
this definition depends only on the background and the motion of the observer, 
 and not on the choice of coordinates 
or of gauge. It is equivalent to defining:
\begin{align} \clh^+(\tau_0) & \mbox{ is the span of the positive spectrum of } 
	\hat{H}_1(\tau_0) \notag \\
\clh^-(\tau_0) & \mbox{ is the span of the negative spectrum of }
	\hat{H}_1(\tau_0) \notag \end{align}

We can now define $\hat{N}_1(\tau) = \hat{P}^+(\tau) - \hat{P}^-(\tau)$, as in 
Section 3.3. We define Bogoliubov coefficients $\balpha(\tau_1,\tau_0), 
\bbeta(\tau_1,\tau_0) \dots$, S-Matrix elements \linebreak $\la \mbox{$\binom{i'_1 i'_2 \dots 
i'_{m'}}{j'_1 j'_2 \cdots j'_{n'}}$}_{\tau_1} | \mbox{$\binom{i_1 i_2 \dots 
i_m}{j_1 j_2 \dots j_n}$}_{\tau_0}(\tau_1) \ra$ and so on entirely as before, and all formulae 
of the previous Sections still apply with $\tau$ in place of $t$, and 
`on $\Sigma_{\tau}$' in place of `at time $t$'. This construction is easily 
generalised to encompass gravitational backgrounds; see \cite{Me2,mythesis}. In 
the next Section it is shown 
that, in the conventional approach to background QFT, this definition corresponds 
to Hamiltonian diagonalisation of the second quantized Hamiltonian obtained by 
substituting the field operator $\hat{\psi}(x)$ into expression (\ref{eq:Hsig}) for $H_{\tau_0}(\psi)$. This definition  
can be seen as a generalisation of Gibbons' approach \cite{Gibb2} to 
arbitrary observers.

It is clear from the definition of radar time that $\Sigma_{\tau_1}$ lies to the 
future of $\Sigma_{\tau_0}$ (for $\tau_1 > \tau_0$) except at the observer's 
particle horizon (supposing one exists), at which point the various $\Sigma_{\tau}$ 
converge. The domain of $\tau(x)$ is not necessarily all 
of spacetime - only that part with which the observer can 
communicate (i.e. send and receive signals). The simplest illustration of this fact involves  
a uniformly accelerating observer.

\begin{figure}[h]
\center{\epsfig{figure=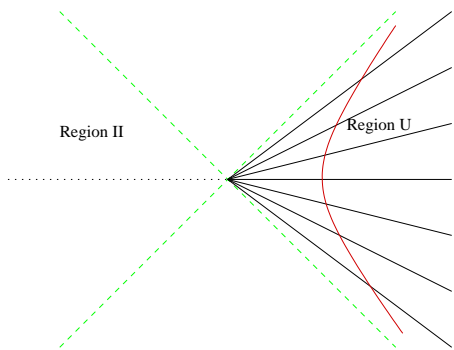, width=6cm}}
\caption{{\footnotesize Hypersurfaces of simultaneity of a uniformly accelerating observer.}}
\end{figure}

For a uniformly accelerating observer, $\rho(x) = \frac{1}{2a} \log(a^2 (z^2 - t^2))$ and 
$\tau(x) = \frac{1}{2a} \log(\frac{z + t}{z - t})$, which are simply 
Rindler coordinates, covering only the region U in Figure 2. Also, 
$k = z \frac{\partial}{\partial t} - t \frac{\partial}{\partial z}$ (or 
$ = \frac{\partial}{\partial \tau}$ in Rindler coordinates), which is the 
Killing vector field used to define positive/negative frequency 
modes in conventional derivations of the Unruh effect. For the 
significance of the dotted line in region II, consider the `Finite 
Acceleration Time' case shown in Figure 3. 

\begin{figure}[h]
\center{\epsfig{figure=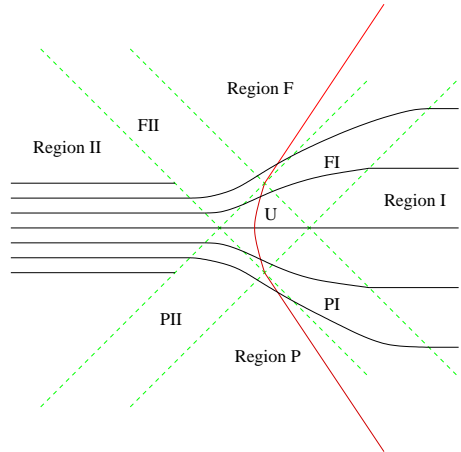, width=6cm}}
\caption{{\footnotesize Hypersurfaces of simultaneity of an observer undergoing uniform acceleration for a finite period of time (in Region U here) who is otherwise inertial.}}
\end{figure}

In the 
limit as this `Acceleration Time' approaches infinity, this case approaches that of a uniformly accelerating observer. In this limit the hypersurfaces of simultaneity 
(which are all Cauchy) all approach this dotted line in region II, and 
$k_{\mu}(x) \rightarrow 0$ there. A particle horizon forms as shown at the origin of Figure 2, and $\clh^0$ is no longer empty but 
rather consists of any states which are zero throughout region U. These state 
are `unmeasurable' by this accelerating observer, and must be traced out in derivations 
of the Unruh effect. This is 
explained in further detail in \cite{Me4}, and Figure 3 is also discussed in \cite{Me3}.

\subsection{Nonlocality of The Particle Definition}

The particle definition presented in this paper, along with the 
definition of vacuum that accompanies it, are nonlocal in  
that they depend on the observer's foliation of spacetime 
into `space at time $\tau$'. We can easily 
see this since declaring a system to 
be `vacuum at time $\tau$' involves asserting that ``at time $\tau$ there were no 
particles anywhere''. This statement requires knowledge of a whole 
spacelike hypersurface defining space `at time $\tau$'. We might have hoped, 
however, to define a 
local `particle density' $N(x)$ with respect to which the 
`vacuum at time $\tau$' could be defined by requiring 
that $N(x|_{\Sigma_{\tau}}) = 0$ in that state. But this too is impossible, 
since no local particle density can be consistent with the 
Unruh effect. 

 To investigate this nonlocality, consider the simple case of the number operator 
at time $\tau$ for an inertial observer in Minkowski space. By working in that 
observer's rest frame we identify $\tau = t$, and we can write \cite{Guth}:
\begin{align} \hat{N}_1 \psi(\bx,t) & \equiv \hat{P}^+ \psi(\bx,t) - 
\hat{P}^- \psi(\bx,t) \\
 & = \int {\rm d}^3 \by \ G(\bx - \by) \psi(\by,t) \label{guth} \\
\mbox{where } G(\bx - \by) & = \int \frac{{\rm d}^3 \bp}{(2 \pi)^3} e^{i \bp \cdot (\bx - \by)}
\frac{\overline{\bp} + m \gamma^0}{\sqrt{|\bp|^2 + m^2}} \end{align}
where $\overline{\bp} \equiv \sum_{k=1}^{3} p^k \gamma_k \gamma_0$. To derive this we simply expand an arbitrary state $\psi(\bx)$ in terms of 
positive and negative energy plane wave states and substitute into (\ref{guth}). 
(The physical extension of this can be written in terms of a field operator as 
$\hat{N} = : \int {\rm d}^3 \bx {\rm d}^3 \by \hat{\psi}(\bx) G(\bx - \by) 
\hat{\psi}(\by) :$ and is the standard number operator used for free Dirac field 
theory.) 
It is straightforward to calculate $G(\bx - \by)$, and we get

$$ G(\bx) = \frac{m^2}{2 \pi^2 |\bx|} \left( K_1(m |\bx|) \gamma^0 + i K_2 (m |\bx|) \frac{\overline{\bx}}{|\bx|} \right) $$

 where $K_n(m |\bx|)$ are Bessel functions, which fall off exponentially 
for $| \bx | >> \lambda_c$. Hence, as we suspected, the 
particle definition is nonlocal on small scales, with the 
nonlocal contribution becoming negligible on scales larger than 
$\lambda_c$. This nonlocality need cause no concern, since it can be 
related directly to the motion of the particle detector. Indeed, the appearance at the heart of the interpretation of QFT of the observer and their
	hypersurface of simultaneity is quite natural. An ability to consider nonlocal
	issues within
	a relativistic context clearly requires (as we
	have done) that a choice of hypersurface be associated with
	any measurement, while consistency requires that this hypersurface be 
specified in terms of the motion of an observer. 	

 	On the other hand, the equation governing the evolution 
of states (given by (\ref{eq:diss2.2}) and (\ref{eq:evol})) remains 
local. The nonlocality 
discussed in this Section relates only to the calculation of expectation 
values in these states, and does not affect their evolution.

\section{RELATIONSHIP TO CONVENTIONAL METHODS}

	In previous Sections we have shown that QFT in electromagnetic 
backgrounds can be performed with no mention of the `field operator' 
$\hat{\psi}(\bx)$, provided we use a suitable representation of the states 
involved. But if we choose, we can for instance define the field operator just as 
in multiparticle quantum mechanics \cite{Szab,MYS}:
\begin{equation} \hat{\psi}(\bx) = \sum_i \psi_{i}(\bx) i_{\psi_{i}} \label{eq:fieldop} \end{equation}
	where the $\{ \psi_{i}(\bx) \}$ form an arbitrary orthonormal 
basis for $\clh$. 
This automatically satisfies the Canonical Anticommutation Relations in 
view of the definition of $i_{\psi}$. It is also independent 
of the choice of basis, so that it does not depend on the 
split of solution space into `$+ve/-ve$ energy solutions'. Operators 
$\hat{A}_H:\clf_{\wedge}(\clh) \rightarrow \clf_{\wedge}(\clh)$ can then be 
written as bilinears of the field operator, and the entire Canonical `second 
quantization' procedure can be set up.

However, in conventional methods we do not have a concrete 
representation of state space, so that we cannot {\it derive} properties 
of the field operator (such as the Canonical Anticommutation Relations). 
Rather we must {\it require} these properties, and then use them to 
deduce the properties of the creation and annihilation operators and 
hence of the state space. In this Section we review this Canonical 
construction, and compare the conventional derivation of S-Matrix elements 
and expectation values with the derivations given in Section 3. Our
	presentation will follow that used (for real scalar
	fields) in Section 6 of DeWitt \cite{DeWitt}. Interestingly, although the 
scalar field S-Matrix elements 
presented in \cite{DeWitt} have been studied and applied extensively 
since their publication, we believe that the derivation presented in Section 5.2  
represents the first treatment of
	fermionic systems along these lines. Although this derivation is new, it is in keeping with the conventional `canonical' approach, and is
the simplest available `conventional' derivation. These results were also obtained, by very
	different methods, by Schwinger \cite{Sch2}, while some special cases have also been derived by various other 
	authors \cite{GaGi,CoMo1,Dam,Man,Park2}. The equivalent derivation 
for charged bosonic systems is presented in \cite{Me5}.

\subsection{Preliminaries}

The field operator $\hat{\psi}(x)$ is now defined (working now in the 
Heisenberg picture) by
\begin{equation} \hat{\psi}(x) = \sum_{i}  \{ u_{i,\tau_0}(x) a_{i,\tau_0,h} + 
v_{i,\tau_0}(x) b^{\dagger}_{i,\tau_0,h} \} \label{eq:psihatt0} \end{equation}
	where the $u_{i,\tau_0}(x)$ are `states which at time $\tau_0$ 
represent particles', while $v_{i,\tau_0}(x)$ are `states which at 
time $\tau_0$ represent antiparticles'; the subscript $h$ denotes that these 
operators annihilate or create Heisenberg-picture states. Such an interpretation 
was previously only possible in general in the asymptotic limit $\tau_0 
\rightarrow \pm \infty$, and sometimes a particle interpretation was 
completely abandoned \cite{Wa} and the $u_{i,\tau_0}(x)$,  $v_{i,\tau_0}(x)$ 
chosen arbitrarily. Now, of course, we will require that the  
$u_{i,\tau_0}(x)$,  $v_{i,\tau_0}(x)$ be defined as in the 
previous Section. 

	Since we can no longer define creation and annihilations operators, by equations such as (\ref{eq:SSB11}) and (\ref{eq:SSB12}), we must deduce their properties 
from the properties of $\hat{\psi}(x)$. By requiring that $\hat{\psi}(x)$ 
satisfy the CAR's 
we can deduce the standard anticommutation relations between the 
creation and annihilation operators. Next, by requiring that $\hat{\psi}(x)$ be 
independent of $\tau_0$ we can deduce relations such as 
\begin{align} a_{i,\tau_1,h} & = \sum_j \{ \alpha_{i j} a_{j,\tau_0,h} + 
\beta_{i j} b^{\dagger}_{j,\tau_0,h} \} \label{eq:new1} \\
 b_{i,\tau_1,h} & = 
\sum_j \{ \gamma_{i j}^* a^{\dagger}_{j,\tau_0,h} +
\epsilon_{i j}^* b_{j,\tau_0,h} \} \label{eq:new2} \end{align}
 which are the Heisenberg picture equivalents of equations such as (\ref{eq:SSB48new}). 

The (Heisenberg-picture) `in' vacuum 
$| {\rm vac}_{\tau_0},h\ra$ is then defined implicitly by the requirement that
\begin{equation} a_{i,\tau_0,h} | {\rm vac}_{\tau_0},h\ra = 0 = 
b_{i,\tau_0,h} | {\rm vac}_{\tau_0},h\ra 
\hs{1} \la {\rm vac}_{\tau_0},h | {\rm vac}_{\tau_0},h\ra = 1 \label{eq:diss3.6}\end{equation}
	and the `in Fock space' is defined by acting on $| {\rm vac}_{\tau_0},h\ra$ 
with all possible combinations of creation operators, and taking the span of 
the result. The `out' vacuum and the `out Fock space' are defined similarly. Since 
these are Heisenberg picture states it follows that $| {\rm vac}_{\tau_0},h\ra$  represents 
not just $| {\rm vac}_{\tau_0} \ra$, 
	but represents $| {\rm vac}_{\tau_0}(\tau) \ra$ for all $\tau$, so
	that for instance $\la {\rm vac}_{\tau_1} | {\rm vac}_{\tau_0} \ra \lra \la
	{\rm vac}_{\tau_1}(\tau) | {\rm vac}_{\tau_0}(\tau) \ra$, which is independent
	of $\tau$ in view of conservation of the inner product. 

	We now show 
that equations (\ref{eq:psihatt0}) - (\ref{eq:diss3.6}) suffice for us to
	deduce the general S-Matrix element and expectation value of the theory without 
requiring a more concrete representation of the states involved. However 
	these derivations are more difficult, and conceptually more
	obscure, than those given in Section 3.

\subsection{S-Matrix Elements}

Define the arbitrary `in' and `out' states:
\begin{align} | \mbox{$\binom{i_1 i_2 \dots i_m}{j_1 j_2 \dots j_n}$}_{\tau_0},h \ra & \equiv a^{\dagger}_{i_1,\tau_0,h} \dots a^{\dagger}_{i_m,\tau_0,h} 
b^{\dagger}_{j_n,\tau_0,h} \dots b^{\dagger}_{j_1,\tau_0,h} | {\rm vac}_{\tau_0},h \ra \notag \\
\mbox{ and } | \mbox{$\binom{i_1 i_2 \dots i_m}{j_1 j_2 \dots j_n}$}_{\tau_1},h \ra & \equiv a^{\dagger}_{i_1,\tau_1,h} \dots
	a^{\dagger}_{i_m,\tau_1,h} b^{\dagger}_{j_n,\tau_1,h} \dots
	b^{\dagger}_{j_1,\tau_1,h} | {\rm vac}_{\tau_1},h \ra \notag \end{align} We
	wish to calculate $\la \mbox{$\binom{i'_1 i'_2 \dots
	i'_{m'}}{j'_1 j'_2 \dots j'_{n'}}$}_{\tau_1},h
	|\mbox{$\binom{i_1 i_2 \dots i_m}{j_1 j_2 \dots j_n}$}_{\tau_0},h \ra$ using only the formalism set up in this Section. To do
	this, begin by defining:
\begin{align} c_v & \equiv \la {\rm vac}_{\tau_1},h| {\rm vac}_{\tau_0},h\ra \\
	V\mbox{$\binom{i_1 i_2 \dots i_m}{j_1 j_2 \dots j_n}$} &
 \equiv \frac{1}{c_v} \la \mbox{$\binom{i_1 i_2 \dots i_m}{j_1 j_2
 \dots j_n}$}_{\tau_1},h | {\rm vac}_{\tau_0},h \ra \\ \Lambda\mbox{$\binom{i_1
 i_2 \dots i_m}{j_1 j_2 \dots j_n}$} & \equiv \frac{1}{c_v} \la {\rm vac}_{\tau_1},h
 |\mbox{$\binom{i_1 i_2 \dots i_m}{j_1 j_2 \dots j_n}$}_{\tau_0},h \ra \end{align} 

	From the definitions of $V$ and $\Lambda$ we have:
\begin{align} | {\rm vac}_{\tau_0},h \ra  & = c_v \sum_{n, m = 0}^{\infty} \sum_{
\stackrel{\hs{.2} i_1 < i_2 \dots < i_m}{\hs{.2} j_1 < j_2 \dots <j_n}} 
\hs{-.3}
V\mbox{$\binom{i_1 i_2 \dots i_m}{j_1 j_2 \dots j_n}$} 
| \mbox{$\binom{i_1 i_2 \dots i_m}{j_1 j_2 \dots j_n}$}_{\tau_1},h \ra 
\label{eq:diss3.8}\\
| {\rm vac}_{\tau_1},h \ra & = c_v \sum_{n, m = 0}^{\infty} \sum_{
\stackrel{\hs{.2} i_1 < i_2 \dots < i_m}{\hs{.2} j_1 < j_2 \dots <j_n}} 
\hs{-.3} 
\Lambda^{*}\mbox{$\binom{i_1 i_2 \dots i_m}{j_1 j_2 \dots j_n}$}
| \mbox{$\binom{i_1 i_2 \dots i_m}{j_1 j_2 \dots j_n}$}_{\tau_0},h \ra 
\label{eq:diss3.9}\end{align}

	Acting on (\ref{eq:diss3.9}) with $a_{i,\tau_1,h}$ and using
	(\ref{eq:new1}) gives:
\begin{gather} 0 =  \sum_{n, m = 0}^{\infty} \sum_{\stackrel{\hs{.2} 
i_1 < i_2 \dots < i_m}{\hs{.2} j_1 < j_2 \dots <j_n}}
\{ \sum_{r = 1}^{m} (-)^{r-1} \alpha_{i i_r}
 \Lambda^{*}\mbox{$\binom{i_1 i_2 \dots i_m}{j_1 j_2 \dots j_n}$}
 |\mbox{$\binom{i_1 \dots \check{i}_r \dots i_m}{j_1 j_2 \dots j_n}$}_{\tau_0},h \ra \notag \\ 
+ \sum_{j} (-)^m \beta_{i j}
 \Lambda^{*}\mbox{$\binom{i_1 i_2 \dots i_m}{j_1 j_2 \dots j_n}$}
 |\mbox{$\binom{i_1 i_2\dots i_m}{j_1 j_2 \dots j_n j}$}_{\tau_0},h
 \ra  \} \label{eq:diss3.10}\end{gather} 
valid for all $i$. Consideration of 
 the coefficient of $|\mbox{$\binom{ }{j_1}$}_{\tau_0},h \ra$ in
 (\ref{eq:diss3.10}) gives us
\begin{equation} 0 = \sum_{i_1} \alpha_{i i_1} \Lambda^*\mbox{$\binom{i_1}{j_1}$} + 
\beta_{i j_1} \Lambda^*() \notag \end{equation} where $\Lambda() = 
\frac{c_v}{c_v} = 1$. So, if we define the matrix $\bLambda \equiv 
\bepsilon^{-1} \bgamma$ as in (\ref{eq:SSBScase2}), so that $\bLambda^{\dagger} = - \balpha^{-1} \bbeta$ 
(from (\ref{eq:SSB43})), then we have $\Lambda\mbox{$\binom{i_1}{j_1}$} = 
\Lambda_{j_1 i_1}$. By considering the 
coefficient of $|\mbox{$\binom{i_1 i_2 \dots  i_{m-1}}{j_1 j_2 \dots j_n}$}_{\tau_0},h \ra$ in (\ref{eq:diss3.10}) it follows after some work that
\begin{equation} \sum_{l} \alpha_{i l}\Lambda^{*}\mbox{$\binom{i_1 \dots i_{m-1} 
l}{j_1 j_2 \dots j_n}$} = (-)^{n+m} \sum_{r=1}^{n} (-)^r \beta_{i j_r} \Lambda^{*}
\mbox{$\binom{i_1 i_2 \dots i_{m-1}}{j_1 \dots \check{j}_r \dots j_n}$}
\notag \end{equation} Upon multiplying through by $\alpha^{-1}_{i_m i}$,
	summing over $i$ and conjugating, we have
\begin{equation} \Lambda\mbox{$\binom{i_1 i_2 \dots i_m}{j_1 j_2 \dots j_n}$}
 = (-)^{n+m} \sum_{r=1}^{n} (-)^{r+1} \Lambda_{j_r i_m}
	\Lambda\mbox{$\binom{i_1 i_2 \dots i_{m-1}}{j_1 \dots
	\check{j}_r \dots j_n}$} \label{eq:diss3.11} \end{equation}
	For $n=m$ this gives (by induction)

\begin{equation} \Lambda\mbox{$\binom{i_1 i_2 \dots i_n}{j_1 j_2 \dots j_n}$} = 
(-)^{\frac{n}{2}(n-1)} \det \left[
 \begin{array}{ccc}  \Lambda_{j_1 i_1} & \cdots & \Lambda_{j_1 i_{n}}
 \\ \vdots & & \vdots \\ \Lambda_{j_{n} i_1} & \cdots & \Lambda_{j_{n}
 i_{n}} \end{array} \right] \label{eq:Smat1} \end{equation} 
in agreement with (\ref{eq:SSBScase2}). By considering the coefficient of
 $|\mbox{$\binom{i_1 \dots i_{m-1}}{ }$}_{\tau_0},h \ra$ in (\ref{eq:diss3.10})
 we have $\Lambda\mbox{$\binom{i_1 i_2 \dots i_m}{ }$} = 0$ whenever $m
 >0$, while it can be deduced that $\Lambda\mbox{$\binom{ }{j_1 j_2 \dots j_n}$} = 0$ 
 by acting on (\ref{eq:diss3.9}) with $b_{i,t_1,h}$ and
 looking at the coefficient of $|\mbox{$\binom{ }{j_1 j_2 \dots
 j_{n-1}}$}_{\tau_0},h \ra$. Combined with (\ref{eq:diss3.11}), this gives 
$\Lambda\mbox{$\binom{i_1 i_2 \dots i_m}{j_1 j_2 \dots j_n}$} =
0$  for  $m \neq n$, as required for agreement with (\ref{eq:SSBScase2}). Similarly, by acting 
on (\ref{eq:diss3.8}) with $a_{i,\tau_0,h}$ and $b_{i,\tau_0,h}$ and extracting 
components we can deduce that
\begin{equation} V\mbox{$\binom{i_1 i_2 \dots i_m}{j_1 j_2 \dots j_n}$} = 
\delta_{m n} (-)^{\frac{n}{2}(n-1)} \det \left[
 \begin{array}{ccc}  V_{i_1 j_1} & \cdots & V_{i_1 j_{n}}
 \\ \vdots & & \vdots \\ V_{i_{n} j_1} & \cdots & V_{i_{n}
 j_{n}} \end{array} \right] \label{eq:Smat2} \end{equation} 
	To complete the rederivation of (\ref{eq:SSBScase1}) and (\ref{eq:SSBScase2})	
we now calculate $c_v$, using 
\begin{align} 1 & = \sum_{n, m = 0}^{\infty} \sum_{\stackrel{\hs{.2} 
i_1 < i_2 \dots < i_m}{\hs{.2} j_1 < j_2 \dots <j_n}} \hs{-.3} 
|\mbox{$\binom{i_1 i_2 \dots i_m}{j_1 j_2 \dots j_n}$}_{\tau_1},h \ra \la 
\mbox{$\binom{i_1 i_2 \dots i_m}{j_1 j_2 \dots j_n}$}_{\tau_1},h | \notag \\
	\mbox{to get } \hs{1.8} 1 & = \sum_{n = 0}^{\infty} \sum_{\stackrel{\hs{.2} 
i_1 < i_2 \dots < i_n}{\hs{.2} j_1 < j_2 \dots <j_n}} \hs{-.3} 
| \la \mbox{$\binom{i_1 i_2 \dots i_n}{j_1 j_2 \dots j_n}$}_{\tau_1},h | {\rm vac}_{\tau_0},h \ra |^2 \notag \\
& = |c_v|^2 (1 + \sum_{n = 1}^{\infty} \sum_{\stackrel{\hs{.2} 
i_1 < i_2 \dots < i_n}{\hs{.2} j_1 < j_2 \dots <j_n}} \hs{-.3}
 | \det \left[
 \begin{array}{ccc}  V_{i_1 j_1} & \cdots & V_{i_1 j_{n}}
 \\ \vdots & & \vdots \\ V_{i_{n} j_1} & \cdots & V_{i_{n}
 j_{n}} \end{array} \right] |^2) \notag \\
 & = |c_v|^2 \det(1 + \bV \bV^{\dagger}) \end{align}
	where we have used the matrix identity:
\begin{equation}\det(1 + \bV \bV^{\dagger}) = 1 + \sum_{n = 1}^{\infty} 
\sum_{\stackrel{\hs{.2} i_1 < i_2 \dots < i_n}{\hs{.2} j_1 < j_2 
\dots <j_n}} \hs{-.3}
| \det \left[
 \begin{array}{ccc}  V_{i_1 j_1} & \cdots & V_{i_1 j_{n}}
 \\ \vdots & & \vdots \\ V_{i_{n} j_1} & \cdots & V_{i_{n}
 j_{n}} \end{array} \right] |^2  \notag \end{equation} 
	This identity follows from the Cauchy-Binet formula, applied (in the 
notation of Horn and Johnson \cite{HoJo}, pg 22) to $A = [V,I_{N}] = B^{\dagger}$ 
with $\alpha = \beta = \{1,\dots ,N \}$ and $r = N$. Hence
\begin{align} |c_v|^2 & = \det(1 + \bV \bV^{\dagger})^{-1} \\
 & = \det(\balpha \balpha^{\dagger}) \hs{3} \mbox{ from (\ref{eq:SSB43})} \\
 & = |\det(\balpha(\tau_1,\tau_0))|^2 = |\det(\bepsilon(\tau_1,\tau_0))|^2 \end{align}
	where the Bogoliubov conditions have been used in the last step. So 
we can successfully deduce that $|c_v| = |\det(\bepsilon(\tau_1,\tau_0))|$ 
without a concrete representation of the vacuum state. However, this derivation  
is technically much more difficult than that in Section 3.2; it 
 requires (rather than derives) unitarity of the full multiparticle S-Matrix; 
and it obscures the fact that $\det(\bepsilon(\tau_1,\tau_0))$ is the inner 
product between `in' and `out' Dirac Seas.

\subsection{Expectation Values and Vacuum Subtraction}

Consider now the relation between 
 Hermitian extension and conventional `second quantization'. We  
consider a `first quantized' 
operator $\hat{A}_1(\tau):\clh \rightarrow \clh$, which in general is a 
differential operator on $\Sigma_{\tau}$. By inserting factors 
of $1 = \sum_i | u_{i, \tau_0}(\tau) \ra \la u_{i, \tau_0}(\tau) | + | v_{i, \tau_0}(\tau) \ra \la v_{i, \tau_0}(\tau) |$ to either side of $\hat{A}_1(\tau)$ we can write:
\begin{align} \hat{A}_1(\tau) & = \sum_{i j} \la u_{i, \tau_0}(\tau) | \hat{A}_1(\tau) 
| u_{j, \tau_0}(\tau) \ra |u_{i, \tau_0}(\tau) \ra \la u_{j, \tau_0}(\tau) | \notag \\
& +  \la u_{i, \tau_0}(\tau) | \hat{A}_1(\tau) | v_{j, \tau_0}(\tau) \ra |u_{i, \tau_0}(\tau) \ra \la v_{j, \tau_0}(\tau) | \notag \\
& + \la v_{i, \tau_0}(\tau) | \hat{A}_1(\tau) | u_{j, \tau_0}(\tau) \ra |v_{i, \tau_0}(\tau) \ra \la u_{j, \tau_0}(\tau) | \notag \\
& + \la v_{i, \tau_0}(\tau) | \hat{A}_1(\tau) | v_{j, \tau_0}(\tau) \ra |v_{i, \tau_0}(\tau) \ra \la v_{j, \tau_0}(\tau) | \notag \end{align}
	Since under Hermitian extension 
$|u_{i, \tau_0}(\tau) \ra \la u_{j, \tau_0}(\tau) | \rightarrow 
a^{\dagger}_{i,\tau_0}(\tau) a_{j,\tau_0}(\tau)$, \linebreak $|u_{i, \tau_0}(\tau) \ra 
\la v_{j, \tau_0}(\tau) | \rightarrow a^{\dagger}_{i,\tau_0}(\tau) 
b^{\dagger}_{j,\tau_0}(\tau)$ etc.. (in the notation introduced in
 Section 3.2), we can now write $\hat{A}_H(\tau)$ as:
\begin{align} \hat{A}_H(\tau) & = \sum_{i j} \{ \la u_{i, \tau_0}(\tau) 
| \hat{A}_1(\tau) | u_{j, \tau_0}(\tau) \ra a^{\dagger}_{i,\tau_0}(\tau) 
a_{j,\tau_0}(\tau) \notag \\
 & + \la u_{i, \tau_0}(\tau) | \hat{A}_1(\tau) | 
v_{j, \tau_0}(\tau) \ra a^{\dagger}_{i,\tau_0}(\tau) 
b^{\dagger}_{j,\tau_0}(\tau) \notag \\
 & + \la v_{i, \tau_0}(\tau) | \hat{A}_1(\tau) | u_{j \tau_0}(\tau) 
\ra b_{i,\tau_0}(\tau) a_{j,\tau_0}(\tau) \notag \\
 & + \la v_{i, \tau_0}(\tau) | 
\hat{A}_1(\tau) | v_{j, \tau_0}(\tau) \ra b_{i,\tau_0}(\tau) 
b^{\dagger}_{j,\tau_0}(\tau) \} \label{eq:opmy} \end{align}
	In the conventional approach to second quantization the 
pre-normal-ordered `second quantized' operator $\hat{A}_{naive}(\tau)$ is 
obtained by substituting the field operator $\hat{\psi}(x)$ from (\ref{eq:psihatt0}) 
into the integral expression for the `first quantized expectation 
value' 
$\la \psi(\tau) | \hat{A}_1(\tau) | \psi(\tau) \ra$. This gives:
\begin{align} \hat{A}_{{\rm naive}}(\tau) & = \sum_{i j} \la u_{i, \tau_0}(\tau) 
| \hat{A}_1(\tau) | u_{j, \tau_0}(\tau) \ra a^{\dagger}_{i,\tau_0,h}
a_{j,\tau_0,h} \notag \\
 & + \la u_{i, \tau_0}(\tau) | \hat{A}_1(\tau) | 
v_{j, \tau_0}(\tau) \ra a^{\dagger}_{i,\tau_0,h} 
b^{\dagger}_{j,\tau_0,h} \notag \\
 & + \la v_{i, \tau_0}(\tau) | \hat{A}_1(\tau) | u_{j \tau_0}(\tau) 
\ra b_{i,\tau_0,h} a_{j,\tau_0,h} \notag \\
 & + \la v_{i, \tau_0}(\tau) | 
\hat{A}_1(\tau) | v_{j, \tau_0}(\tau) \ra b_{i,\tau_0,h} 
b^{\dagger}_{j,\tau_0,h} \label{eq:opconven} \end{align} which is
	clearly the Heisenberg picture version of
	(\ref{eq:opmy}). By construction, neither of these
	operators depends on the choice of $\tau_0$. It is also worth
	noting that although (\ref{eq:opmy}) and (\ref{eq:opconven})
	are equivalent, the expression (\ref{eq:opmy}) is seldom very
	useful in our formalism. It is usually more convenient 
	to work directly from the definition
	(\ref{eq:SSB8}). Likewise, in conventional
	multiparticle quantum mechanics (see \cite{Szab} for instance)
	expressions like (\ref{eq:opmy}) appear (with only the
	$u_{i}(x)'s$ and $a_{i}$'s), but are seldom useful. This 
	is partly because writing down (\ref{eq:opmy})
	demands a complete set of solutions
	to the (first quantized) governing equation; the
	usefulness of (\ref{eq:opmy}) then relies on the matrix
	elements taking a convenient form in this set. Indeed, even
	writing down the field operator $\hat{\psi}(x)$ requires this
	complete set of solutions. By contrast, the definition
	(\ref{eq:SSB8}), which is precisely the
	definition used in multiparticle quantum mechanics, is
	well-defined whether or not we have such a
	complete set. \newline

We can now calculate the expectation value of  $\hat{A}_{{\rm naive}}(\tau)$ 
in the state $| {\rm vac}_{\tau_{in}},h \ra$, representing the state that `at 
time $\tau_{in}$' was vacuum. To do this 
choose $\tau_0 = \tau_{in}$ in (\ref{eq:opconven}), so that
\begin{equation} \la {\rm vac}_{\tau_{in}},h | \hat{A}_{{\rm naive}}(\tau) | {\rm vac}_{\tau_{in}},h \ra 
 = \sum_{i} \la v_{i, \tau_{in}}(\tau) | \hat{A}_1(\tau) 
| v_{i \tau_{in}}(\tau) \ra \end{equation}
	This is simply the first term in (\ref{eq:vacsub}), which followed directly 
from (\ref{eq:prop3a}). Although $\hat{A}_{{\rm naive}}(\tau)$ is independent of the choice of 
$\tau_0$, the normal ordering process clearly depends on $\tau_0$. We denote {\it
normal ordering with respect to the $\tau'$ expansion of
$\hat{\psi}(x)$} by $:\hs{1}:_{\tau'}$.
	Then:
\begin{align} :\hat{A}_{{\rm naive}}(\tau):_{\tau'} & = \sum_{i j} \la u_{i, \tau'}(\tau) 
| \hat{A}_1(\tau) | u_{j, \tau'}(\tau) \ra a^{\dagger}_{i,\tau',h}
a_{j,\tau',h} \notag \\
 & + \la u_{i, \tau'}(\tau) | \hat{A}_1(\tau) | 
v_{j, \tau'}(\tau) \ra 
a^{\dagger}_{i,\tau',h} b^{\dagger}_{j,\tau',h} \notag \\
 & + \la v_{i, \tau'}(\tau) | \hat{A}_1(\tau) | u_{j \tau'}(\tau) 
\ra b_{i,\tau',h} a_{j,\tau',h} \notag \\
& - \la v_{i, \tau'}(\tau) | 
\hat{A}_1(\tau) | v_{j, \tau'}(\tau) \ra b^{\dagger}_{j,\tau',h} 
b_{i,\tau',h} \label{eq:normord1} \\
 & = \hat{A}_{{\rm naive}}(\tau) - \sum_{i} \la v_{i, \tau'}(\tau) | 
\hat{A}_1(\tau) | v_{i, \tau'}(\tau) \ra \hat{1} \label{eq:normord2} \end{align}

	Comparison of (\ref{eq:normord2}) with (\ref{eq:vacsub}) 
reveals that vacuum subtraction corresponds to the choice $\tau' = \tau$, 
and so to normal ordering {\it at the time of measurement}:
\begin{equation} \hat{A}_{{\rm phys}}(\tau) \hs{.2} \Leftrightarrow  \hs{.2} 
:\hat{A}_{{\rm naive}}(\tau):_{\tau} \end{equation}
	This is also the choice used in the `Hamiltonian diagonalisation' procedure 
discussed in the next Section. From now on we 
shall assume that $\tau' = \tau$. Expression 
(\ref{eq:normord1}) is obtained upon normal ordering by `moving the creation operators 
to the left, and changing appropriate signs', while (\ref{eq:normord2}) 
acknowledges that this is the same as subtracting an appropriate 
multiple of the unit operator. Since 
(\ref{eq:normord2}) still allows us to express $\hat{A}_{{\rm naive}}(\tau)$ in 
terms of an arbitrary $\tau_0$ it is generally more convenient 
than (\ref{eq:normord1}), which involves expressing $\hat{A}_{{\rm naive}}(\tau)$ 
in terms of 
$\tau_0 = \tau' (= \tau$ now). \newline

It will be convenient for later to write the 
operator $\hat{H}_{{\rm naive}}(\tau)$ (obtained by substituting $\hat{\psi}(x)$ 
into (\ref{eq:Hsig})) expressed in terms of $\tau_0 = \tau$. This can be written as:
\begin{gather} \hat{H}_{{\rm naive}}(\tau) = \sum_{i j} \bH^{++}_{i j}(\tau) a^{\dagger}_{i,\tau,h}
a_{j,\tau,h} + \bH^{+-}_{i j}(\tau) a^{\dagger}_{i,\tau,h} 
b^{\dagger}_{j,\tau,h} \notag \\
+ \bH^{-+}_{i j}(\tau) b_{i,\tau,h} a_{j,\tau,h} + \bH^{--}_{i j}(\tau)  b_{i,\tau,h} 
b^{\dagger}_{j,\tau,h} \label{eq:hdiag1} \end{gather}
	Since $\{  u_{j, \tau_0}(\tau) \}$ 
and $\{ v_{j, \tau_0}(\tau) \}$ are chosen to satisfy 
the particle definition introduced in Section 4, we clearly 
have $\bH^{-+}(\tau) = 0 =\bH^{+-}(\tau)$, 
so that $\hat{H}_{{\rm naive}}(\tau)$ contains no 
$a^{\dagger}_{i,\tau,h} b^{\dagger}_{j,\tau,h}$, or $b_{i,\tau,h} a_{j,\tau,h}$ 
terms. It also follows that $\bH^{++}(\tau)$ and $-\bH^{--}(\tau)$ are positive definite. 
Conversely, it is straightforward to show that if $\bH^{-+}(\tau) = 0 
=\bH^{+-}(\tau)$, (and $\bH^{++}(\tau)$ and $-\bH^{--}(\tau)$ are positive 
definite) then the $\{  u_{j, \tau_0}(\tau) \}$ span 
$\clh^{+}(\tau)$, and $\{ v_{j, \tau_0}(\tau) \}$ span $\clh^{-}(\tau)$. 
This verifies that the particle definition given in Section 4 is 
equivalent to Hamiltonian diagonalisation of the Hamiltonian obtained by 
substituting $\hat{\psi}(x)$ into (\ref{eq:Hsig}). We now discuss Hamiltonian diagonalisation.

\subsection{On Hamiltonian Diagonalisation}

The idea of using `Hamiltonian diagonalisation' as a 
prescription for defining particle/antiparticle states dates 
back to Imamura \cite{Imam} in 1960, 
based on an analogy with Bogoliubov's work on superfluidity \cite{Bog1} 
and superconductivity \cite{Bog2}. It was developed 
in more detail by Grib and Mamaev \cite{GM1,GM2} and Grib, 
Mamayev and Mostepanenko \cite{GMM} duting the 1970's. The basic idea is to 
expand a second quantized Hamiltonian $\hat{H}(\tau_0)$ in terms of 
creation/annihilation operators (just as in (\ref{eq:hdiag1})), and choose 
the basis solutions that `represent particle/antiparticle states 
at time $\tau_0$' by requiring that $\hat{H}(\tau_0)$ contain no terms proportional to $b_{i,\tau_0} a_{j,\tau_0}$ or $a^{\dagger}_{i,\tau_0} b^{\dagger}_{j,\tau_0}$.
Application of this requirement at two different times leads to two 
different mode-expansions of $\hat{\psi}(x)$, related by Bogoliubov coefficients, 
which can then be used to calculate S-matrix elements, vacuum expectation values etc.

	This procedure clearly depends on the choice of second quantized Hamiltonian. 
As pointed out by Fulling \cite{Full}, previous prescriptions for 
choosing a Hamiltonian \cite{MMS,GMM,GMM2} were far from unique, since they 
depended on an arbitrarily chosen `time' coordinate. The prescription of 
Gibbons \cite{Gibb2}, which amounts to diagonalising
	$\int T_{\nu \mu} k^{\nu} {\rm d} \Sigma^{\mu}$ where $k^{\mu}$ is
	a timelike Killing vector field, is less arbitrary, since it depends 
only on the choice of timelike Killing vector field. However, this restricts 
us to spacetimes which admit timelike Killing vector fields, and restricts
us to considering observers travelling on the integral curves of those Killing 
vector fields. Our particle
 definition, which amounts to diagonalising the Hamiltonian obtained by 
substituting $\hat{\psi}(x)$ into (\ref{eq:Hsig}), is precisely the definition needed to fix these
 problems. This Hamiltonian, which has not been studied before, is 
the first Hamiltonian that is defined directly in terms of the motion of an observer 
and that depends {\it only} on the choice of observer and the background 
present - not on the choice of coordinates, the choice of gauge or the detailed
	construction of the observer's particle detector. It provides a unique  
particle interpretation for any chosen observer travelling in an arbitrary electromagnetic 
background. Generalisation   
to curved spacetimes is straightforward, and is presented in \cite{Me2}. 

	Another 
advantage of the prescription presented in Sections 3 and 4 
is that it does more than provide a `Hamiltonian
 diagonalisation condition' (equation (\ref{eq:Hdef})) which
 $\clh^{\pm}(\tau_0)$ must satisfy. It also provides a clear
 procedure for constructing $\clh^{\pm}(\tau_0)$ (in terms of
 eigenstates of the 1st quantized Hamiltonian $\hat{H}_1(\tau)$) which
 allows us to tackle arbitrarily complicated situations. With
 conventional Hamiltonian diagonalisation prescriptions unless
 the spacetime allows separation of variables no clue is given 
 how to satisfy the `Hamiltonian diagonalisation
 condition'. This need to separate variables also plagues 
the `adiabatic approximation' techniques
	advocated by Fulling \cite{Full,Full2} and others. 

Another particle definition is that used by Fradkin, Gitman 
and Shvartsman \cite{FGS}. As here, they use eigenstates 
of a `first quantized Hamiltonian' at times $t_{in}$ and
$t_{out}$ to define `in' and `out' states (though they do not consider non-inertial observers). However they use the spectrum of $\hat{H}_{ev}(t)$ rather than
$\hat{H}_1(t)$ making their
formulation gauge dependent! To see this one need
only consider free field theory in the gauge $A(x) = (A,0,0,0)$, where
$A$ is some constant, for which $e A \gg m$. A more telling example
arises in the familiar case of a constant electric field, where if the gauge $A
= (E z ,0,0,0)$ is chosen, then no pair creation  
occurs in Fradkin et al.'s \cite{FGS}  approach! Gauge dependence in the choice of
particle interpretation is a problem that has troubled all previous
`Bogoliubov coefficient' approaches to quantum field
theory in electromagnetic backgrounds \cite{SPad}. The problem is usually `patched up' by
appealing to the tunnelling interpretation. If the gauge $A =
(E z ,0,0,0)$ is chosen, particle creation can be described in terms
of particles `tunnelling through the barrier that separates
particle/antiparticle states' (see \cite{Dam,GMR} or \cite{BMPS} for a
description of these methods). This leaves us with two theories, each
failing where
the other succeeds. Here the two approaches arise
naturally as parts of the same formalism. By using eigenstates of the
gauge-covariant Hamiltonian $\hat{H}_1 \neq \hat{H}_{ev}$ to define
particle states, we see that particle creation in the background
$A^{\mu}(x)$ requires either: (1) that $A^0(x) \neq 0$, so that $A^0(x)$ can act
as a potential, w.r.t. which tunnelling may occur, or (2) 
$\bA(x) = (A^1(x),A^2(x),A^3(x))$ is
time-dependent, so that eigenstates of $\hat{H}_1(t)$ mix during
evolution. Both effects contribute to the calculation
of Bogoliubov coefficients \cite{mythesis}. An immediate consequence is that if a
gauge can be chosen such that $A^0 = 0$ and $\bA$ is time-independent,
then particle creation will not occur. That is, {\it time-independent
magnetic fields cannot create particles}. This result agrees
with Schwinger's calculation \cite{Schw} in terms of an effective
Lagrangian, and successfully resolves all of the inconsistency problems
raised by Sriramkumar and Padmanabhan \cite{SPad}. Another simple example is 
the case of electrostatic fields, where a gauge can be chosen such that $A^0$ 
is time-independent, while $\bA = 0$. In this case the physical vacuum is 
constructed from negative energy solutions of the free Dirac equation, 
and stationary states are constructed from eigenstates 
of $\hat{H}_{{\rm ev}}$. The difference between the physical vacuum and the `lowest 
energy stationary state' accounts for vacuum effects such as Lamb screening.

\section{DISCUSSION}

	We have presented above a formulation of fermionic quantum field
	theory in electromagnetic backgrounds that is precisely analogous to the methods used in multiparticle quantum 
mechanics. The only difference is that the particle interpretation requires us to consider the entire Dirac Sea, as well as any particles which may be present. Rather than working with a field operator $\hat{\psi}(x)$ and an 
abstract Fock space constructed solely from creation and annihilation operators 
acting on some postulated vacuum $| 0 \ra$, we work directly with the Schr\"{o}dinger 
picture states of the system, described in terms of Slater determinants of 
solutions of the Dirac equation. As 
well as providing a
conceptually transparent approach to the theory and an extremely 
simple derivation of the general S-Matrix element and expectation value of the theory, 
this approach also allows us to provide a consistent particle interpretation for all times, without 
requiring any `asymptotic niceness conditions' on the `in' and `out' states. Other
 	advantages include the ease with
 	which unitarity of the S-Matrix follows from conservation of
 	the Dirac inner product,
 	insights into quantum anomalies, and the fact 
that Hermitian extension provides well-defined second quantized operators without 
requiring a complete set of orthonormal modes. We have used the 
concept of `radar time' to  
generalise the particle interpretation to an arbitrarily moving observer, providing a 
definition of particle which depends only on the observer's motion and on the 
background present, not on the choice of coordinates, the choice of gauge, or the 
detailed construction of the particle detector.

\begin{acknowledgments}

	 We thank Dr Jim Reeds, Anton Garrett, Professor Alan Guth and 
Dr Emil Mottola for helpful discussions, Trinity College and Cavendish 
Astrophysics for financial support, and the Institute for Nuclear Theory, 
University of Washington for hospitality during his visit there.

\end{acknowledgments}

\end{article}

\end{document}